\documentclass[twocolumn,reprint,aps,prb, floatfix,amsmath,amssymb,superscriptaddress]{revtex4-2}
\usepackage[utf8]{inputenc}
\usepackage{graphicx}
\usepackage{bm}
\usepackage{braket}
\usepackage{xcolor} 
\usepackage{subfigure}     
\usepackage{amsthm}
\usepackage{hyperref}
\usepackage{makecell}
\newcommand{\CFT}{{\textsl{\tiny CFT}}}
\newcommand{\Tr}{\mathrm{Tr}}
\newcommand{\Pf}{\mathrm{Pf}}
\newcommand{\diag}{\mathrm{diag}}
\newcommand{\cyl}{cyl}

\begin{document}

\title{Multi-boundary generalization of thermofield double states \\
and their realization in critical quantum spin chains}
\date{\today}
\author{Yijian Zou}
\author{Guifre Vidal}
\affiliation{Sandbox@Alphabet, Mountain View, CA 94043, USA}

\begin{abstract}
We propose a multi-boundary generalization of thermofield double states (TFD) of a two-dimensional conformal field theory (CFT) and show, through a conformal map to the complex plane, that they are closely related to multi-point correlation functions. 
We then also describe how to approximately realize these multi-boundary TFD states numerically on the lattice, starting from a critical quantum spin chain Hamiltonian. In addition, finite size corrections on the lattice are seen to be significantly reduced by the use of \textit{smoothers} -- numerically optimized unitary transformations that locally re-arrange the quantum spin degrees of freedom. One merit of the spin chain realization is that it allows us to probe the properties of the proposed multi-boundary TFD states through numerical experiments, including the characterization of their entanglement structure.
As an illustration, we explicitly construct generalized TFD states with three and four boundaries for the Ising CFT and compute entanglement quantities using novel free fermion techniques. We find ranges of parameters where their multipartite entanglement is significant or negligible.
\end{abstract}

\maketitle

Thermofield double (TFD) states play a distinguished role in several areas of quantum physics, including quantum many-body physics, quantum fields, and quantum gravity. They appear naturally e.g. in the study of quantum spin chains and quantum fields at finite temperature, as well as in the holographic description, in terms of a conformal field theory (CFT) of certain class of gravitational wormholes connecting two black holes \cite{Maldacena_2001,Maldacena_2013,Balasubramanian_2014,Chapman_2018,Yang_2018}. Lately, the experimental realization of TFD states has been discussed \cite{Cottrell_2018,Wu_2018,Zhu_2019}, including proposals involving quantum computers. 

Consider a system with Hamiltonian $H$ in a thermal state $\rho_{\beta} = e^{-\beta H}/Z$, where $\beta$ is the inverse temperature and $Z(\beta)= \mbox{tr}(e^{-\beta H})$ is a normalization constant. As further reviewed below, a TFD state $\ket{\psi_{\rho_\beta}^{(1,2)}}$ can be understood as a symmetric purification of the thermal state $\rho_{\beta}$, which is an entangled pure state supported on two identical copies of the system, such that the thermal state in one copy (system 1) is recovered upon tracing out over the second copy (system 2), that is $\rho_{\beta} = \mbox{tr}_{(2)}(\ket{\psi_{\rho_\beta}^{(1,2)}}\bra{\psi_{\rho_\beta}^{(1,2)}})$. A TFD state can also be obtained by regarding a (suitably normalized) imaginary time evolution operator $e^{-(\beta/2)H}/Z(\beta/2)$ as a wavefunction or, alternatively, by a closely related, euclidean path integral on an appropriate geometry. For instance, for a 1+1 dimensional quantum field theory on a circle, a TFD state on two copies of the circle is obtained by an Euclidean path integral on a cylinder. In particular, for a 1+1 CFT on the circle, the above TFD state has been proposed in AdS/CFT to be dual to a 2+1 dimensional quantum gravity geometry corresponding to a two-sided wormhole connecting two black holes \cite{Maldacena_2001,Maldacena_2013,Balasubramanian_2014,Chapman_2018,Yang_2018}.

Generalizations of TFD states connecting more than two copies of the system have been proposed in the context of holography. Specifically, Ref.~\cite{Balasubramanian_2014} investigated the entanglement properties of a family of so-called multi-boundary wormhole states, defined in terms of a CFT Euclidean path integral over a Riemann surface with $n$ holes, which reduce to a TFD state for $n=2$. It was argued that in certain limit the corresponding $n$-partite wavefunction could be written in terms of the n-point functions of the CFT.

In this paper we also propose a family of generalized TFD states for 1+1 CFTs. As in Ref.~\cite{Balasubramanian_2014}, our construction produces entangled states of n copies of a quantum system by means of an Euclidean path integral. However, the geometries with n boundaries that we use as a generalization of a cylinder differ from those in Ref.~\cite{Balasubramanian_2014}. 
We then present a number of results. Firstly, using a conformal map from the n-boundary geometry to the complex plane, we show that the resulting generalized TFD wavefunction encodes $n$-point correlation functions of the CFT. Next, we describe how to build a lattice version of the generalized TFD states, starting from a critical quantum spin chain Hamiltonian. On the lattice, the properties of the TFD states mimic those of the CFT in the continuum, but display strong finite-size corrections. In order to significantly reduce finite-size lattice corrections, we then introduce \textit{smoothers}, numerically optimized unitary transformations that locally re-arrange the quantum spin degrees of freedom. As an illustration of our proposals, we construct the multi-boundary TFD state with three and four boundaries for the Ising CFT. We use novel free fermion techniques that allow us to compute the multi-point correlation functions from the lattice wavefunctions, and compare them to the CFT results.

A first merit of our construction is that, once we have built a generalized TFD state on the lattice, we can numerically study its properties, including its entanglement properties. (This was also possible in \cite{Balasubramanian_2014} for so-called holographic CFTs, whereas our results are valid for any CFT for which we can find a lattice realization). Here we compute bipartite and multi-partite entanglement quantities of the multi-boundary TFD states for the Ising CFT. Furthermore, we identify ranges of parameters where their multipartite entanglement is either significant or negligible. 

A second merit of our proposal is that it allows us to numerically extract the operator product expansion (OPE) coefficients and multi-point correlation functions of a CFT from a critical spin chain that realizes the latter, by computing overlaps between wavefunctions on the lattice. In recent years, related proposals were made that required however computing lattice versions of the CFT primary operators \cite{Zou_2019,Zou_2019_02}. Here, no such lattice scaling operators are needed: the OPE coefficients follows solely from the computation of wavefunction overlaps.

The paper is organized as follows. In Sec.~\ref{sec:TFD} we introduce the multi-boundary generalization of the TFD state and show how the wavefunction is related to the multi-point correlation functions. In Sec.~\ref{sec:lat} we describe how the state can be constructed on the lattice and the use of smoothers to significantly reduce finite-size corrections. In Sec.~\ref{sec:tech} we review the correlation matrix formalism of the free fermion and derive new techniques that are useful for the construction of multi-boundary TFD states. In Sec.~\ref{sec:num} we present numerical results for multi-boundary TFD states for the Ising CFT. We conclude with further discussions in Sec.~\ref{sec:concl}.

\section{Multi-boundary wormhole state}
\label{sec:TFD}
In a CFT, a state can be prepared by a path integral on a manifold with one or more boundaries. In this section we define our proposed class of multi-boundary TFD states for 1+1 CFTs, which can be prepared by a path integral on a strip-like geometry with multiple boundaries. We start with the familiar thermofield double state prepared by the path integral on an open cylinder and then generalize it to three and four-sided TFD states. We then show that multi-point correlation functions of the CFT are encoded in these wavefunctions.
\subsection{Thermofield double state}
The TFD state can be prepared by a Euclidean path integral over an open cylinder. Denote the circumference of the cylinder as $L$ and the length as $\tau_0$. The Hilbert space is supported on the circles that wind around the cylinder. Let $\mathcal{H}^{\CFT}$ be the Hilbert space of the CFT and $H^{\CFT}$ be the Hamiltonian of the CFT. The Hilbert space is spanned by the eigenstates $|\psi_{\alpha}\rangle$ of $H^{\CFT}$, which is in one-to-one correspondence with the scaling operators $\psi_{\alpha}$. The energy of $|\psi_{\alpha}\rangle$ is given by
\begin{equation}
    E_{\alpha} = \frac{2\pi}{L}\left(\Delta_{\alpha}-\frac{c}{12}\right),
\end{equation}
where $\Delta_{\alpha}$ is the scaling dimension and $c$ is the central charge. For example, $\alpha=I$ denotes the ground state of te CFT. Let $T$ be an antiunitary operator (typically chosen as the time reversal operator) that commutes with the Hamiltonian. Then the thermofield double (TFD) state can be defined on a doubled Hilbert space $\mathcal{H}_{12}=\mathcal{H}_1 \otimes \mathcal{H}_2 $ where $\mathcal{H}_1 $ and $\mathcal{H}_2$ are isomorphic to $\mathcal{H}^{\CFT}$,
\begin{equation}
    |\psi(\tau_0)\rangle = \sum_{\alpha} e^{-\tau_0 E_\alpha}|\psi^{1}_\alpha\rangle \otimes |\psi^{2*}_{\alpha}\rangle,
\end{equation}
where we have used the short-hand notation
\begin{equation}
    |\psi^{*}_\alpha\rangle = T|\psi_\alpha\rangle.
\end{equation}
One may interpret the path integral as evolving in imaginary time from a maximal entangled state $|\Omega_{1,2}\rangle$ on $\mathcal{H}_{12}$, 
\begin{equation}
\label{eq:def_TFD}
    |\psi(\tau_0)\rangle = e^{-\tau_0 H_1/2} e^{-\tau_0 H^{*}_2/2}|\Omega_{1,2}\rangle,
\end{equation}
where $H_1 = H \otimes I$ and $H^{*}_2 = I \otimes THT^{-1}$, and 
\begin{equation}
    |\Omega_{1,2}\rangle = \sum_{\alpha} |\psi^{1}_\alpha\rangle \otimes |\psi^{2*}_\alpha\rangle.
\end{equation}
We may write the maximal entangled state in an alternative way
\begin{equation}
    |\Omega_{1,2}\rangle = \sum_{\alpha\beta} A_{\alpha\beta} |\psi^{1}_{\alpha}\rangle \otimes |\psi^{2*}_{\beta}\rangle,
\end{equation}
where
\begin{equation}
    A_{\alpha\beta} = \langle \psi_{\alpha}|\psi_\beta\rangle = \delta_{\alpha\beta}.
\end{equation}
Denoting the wavefunction as
\begin{equation}
    \tilde{A}_{\alpha\beta}(\tau_0) = \langle \psi^{1}_{\alpha} \psi^{2*}_{\beta}|\psi({\tau_0})\rangle,
\end{equation}
we find that the scaling dimensions are encoded in the wavefunction as
\begin{equation}
    \frac{\tilde{A}_{\alpha\beta}(\tau_0)}{{\tilde{A}}_{II}(\tau_0)} = e^{-\frac{2\pi\Delta_{\alpha}\tau_0}{L}}A_{\alpha\beta}.
\end{equation}
\subsection{Three-sided TFD state}
\begin{figure}
    \centering
    \includegraphics[width = \linewidth]{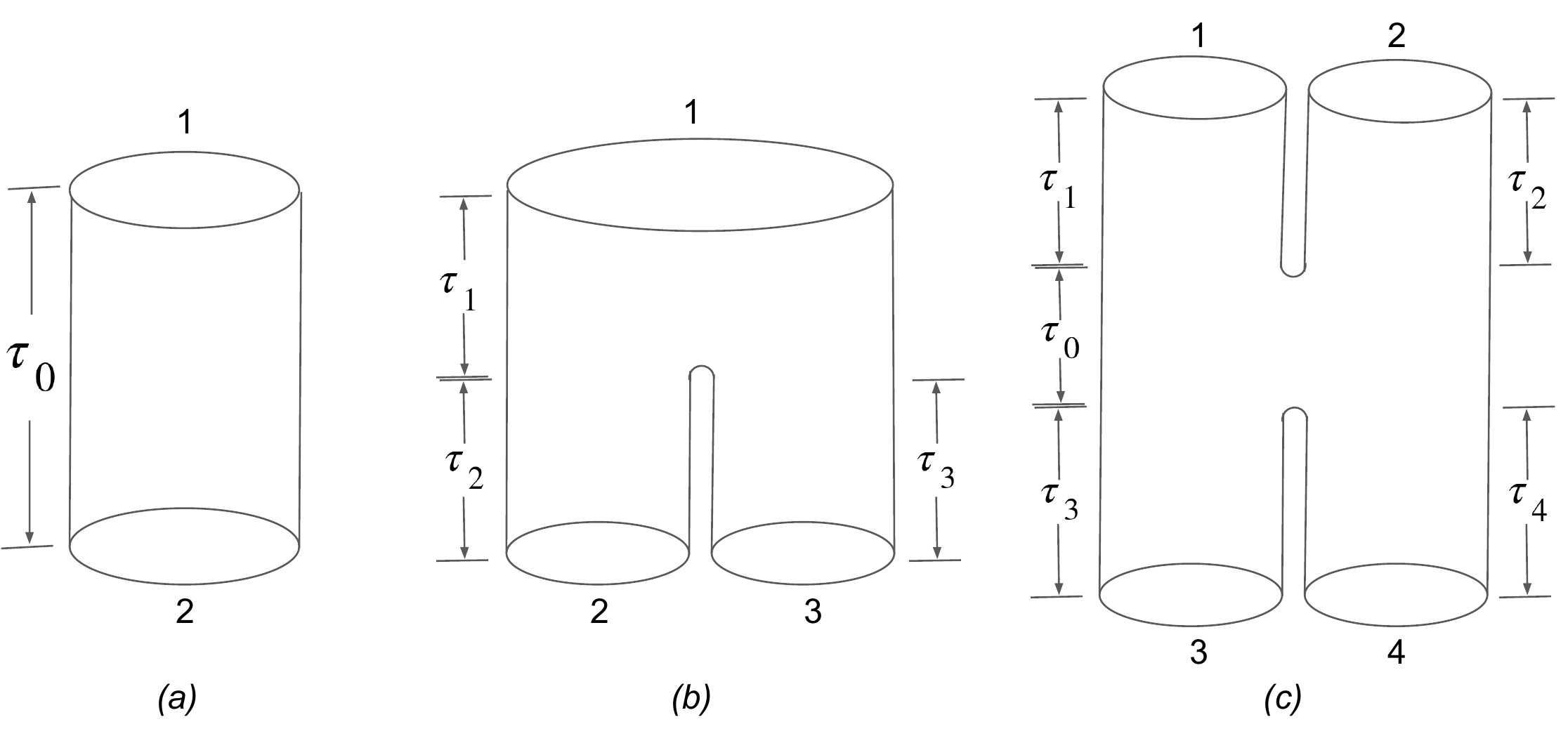}
    \caption{Geometries of the path integrals for multi-boundary TFD states. (a) An open cylinder which corresponds to the TFD state Eq.~\eqref{eq:def_TFD}. (b) The pants-like geometry which corresponds to the three-sided TFD state Eq.~\eqref{eq:def_3sidedBH}. (c) The combination of two pants-like geometry which corresponds to the four-sided TFD state Eq.~\eqref{eq:def_4sidedBH}.}
    \label{fig:geometry}
\end{figure}

Next, we generalize the TFD state to three boundaries. The wave-function can be produced by the CFT path integral over a pants-like geometry (Fig.~\ref{fig:geometry}). The geometry is composed of three cylinders with circumsference $L_1, L_2$ and $L_3$, where $L_1 = L_2 + L_3$. At time $\tau=0$, the geometry encounters a sharp split from one cylinder to two disjoint cylinders. Strictly speaking, the Hilbert space $\mathcal{H}_1$ does not factorize into $\mathcal{H}_2\otimes \mathcal{H}_{3}$ in a quantum field theory. However, as is well known in the study of the entanglement entropy in CFTs, the split is sensible if a UV cutoff is introduced \cite{Calabrese_2004,Calabrese_2009,Cardy_2016}. Alternatively, one may think of the geometry as the limit of a 2D square lattice where the boundary condition changes sharply at $\tau=0$ and the lattice spacing goes to zero (see also Sec.~\ref{sec:lat}). 

The quantum state prepared by the path integral on the pants geometry can be represented by
\begin{equation}
\label{eq:def_3sidedBH}
    |\psi(\tau_1,\tau_2,\tau_3)\rangle = e^{-\tau_1 H_1}e^{-\tau_2 H^{*}_2} e^{-\tau_3 H^{*}_3}|\Omega_{1,23}\rangle,
\end{equation}
where $H_i$ represents the CFT Hamiltonian on the $i$-th circle, and $\tau_i$ is the length of the $i$-th cylinder. The state $|\Omega_{1,23}\rangle$ is an entangled state on $\mathcal{H}_1\otimes\mathcal{H}_{23}$,
\begin{equation}
    |\Omega_{1,23}\rangle = \sum_{\alpha\beta\gamma} A_{\alpha\beta\gamma} |\psi^{1}_{\alpha}\rangle \otimes |\psi^{2*}_{\beta}\rangle \otimes |\psi^{3*}_{\gamma}\rangle
\end{equation}
where 
\begin{equation}
    A_{\alpha\beta\gamma} = \langle \psi^{1}_{\alpha}|\psi^{2}_{\beta}\psi^{3}_{\gamma}\rangle
\end{equation}
Again, the overlap only makes sense if one introduces a UV cutoff (such as a lattice) to the CFT. The coefficients $A_{\alpha\beta\gamma}$ is completely determined by the conformal data. It can be shown, through a singular confomral map that maps the three-sided cylinder to the complex plane, that in the case of $L_1=2L_2=2L_3$ and all indices representing primary states,
\begin{equation}
\label{eq:3_sided_As}
    \frac{A_{\alpha\beta\gamma}}{A_{III}} = 2^{\Delta_{\alpha}-2\Delta_{\beta}-2\Delta_\gamma} C_{\alpha\beta\gamma},
\end{equation}
where $\Delta_{\alpha}$ is the scaling dimension of the primary operator and $C_{\alpha\beta\gamma}$ is the OPE coefficient. The coefficients involving descendant states can also determined by conformal invariance. A detailed proof is presented in the appendix. 

Finally, the three-sided TFD state encodes the conformal data in the following way,
\begin{equation}
\label{eq:3_sided_As_tilde}
    \frac{\tilde{A}_{\alpha\beta\gamma}(\tau_1,\tau_2,\tau_3)}{{\tilde{A}}_{III}(\tau_1,\tau_2,\tau_3)} = e^{-\frac{2\pi\Delta_{\alpha}\tau_1}{L_1}-\frac{2\pi\Delta_{\beta}\tau_2}{L_2}-\frac{2\pi\Delta_{\gamma}\tau_3}{L_3}}A_{\alpha\beta\gamma},
\end{equation}
where
\begin{equation}
    \tilde{A}_{\alpha\beta\gamma}(\tau_1,\tau_2,\tau_3) = \langle \psi^{1}_{\alpha} \psi^{2*}_{\beta} \psi^{3*}_{\gamma}|\psi(\tau_1,\tau_2,\tau_3)\rangle
\end{equation}
is the wavefunction. Below we will compare Eq.~\eqref{eq:3_sided_As} with numerical results on the lattice. We also note that Eq.~\eqref{eq:3_sided_As} as well as Eq.~\eqref{eq:3_sided_As_tilde} leads to a way to compute the OPE coefficients from the wavefunctions,
\begin{eqnarray}
    C_{\alpha\beta\gamma} &=& e^{\frac{2\pi\Delta_{\alpha}\tau_1}{L_1}+\frac{2\pi\Delta_{\beta}\tau_2}{L_2}+\frac{2\pi\Delta_{\gamma}\tau_3}{L_3}} \times \nonumber  \\
    &~&2^{-\Delta_{\alpha}+2\Delta_{\beta}+2\Delta_{\gamma}}\frac{\tilde{A}_{\alpha\beta\gamma}(\tau_1,\tau_2,\tau_3)}{{\tilde{A}}_{III}(\tau_1,\tau_2,\tau_3)}.
\end{eqnarray}
\subsection{Four-sided TFD state}
The four-sided TFD state is defined by the path integral in the geometry shown in Fig.~\ref{fig:geometry} where two pants are connected through the cylinder in the middle with length $\tau_0$. Let the Hamiltonian on the four sides be $H_1,H_2,H_3$ and $H_4$ and the Hamiltonian on the cylinder in the middle be $H_0$. Then the four-sided TFD state is 
\begin{eqnarray}
\label{eq:def_4sidedBH}
    &~&|\psi(\tau_1,\tau_2,\tau_3,\tau_4;\tau_0)\rangle \nonumber \\
    &=& e^{-\tau_1 H_1} e^{-\tau_2 H_2} e^{-\tau_3 H^{*}_3} e^{-\tau_4 H^{*}_4} |\Omega_{12,34}(\tau_0)\rangle,
\end{eqnarray}
where 
\begin{equation}
    |\Omega_{12,34}(\tau_0)\rangle = \sum_{\alpha\beta\gamma\delta} A_{\alpha\beta\gamma\delta}(\tau_0) |\psi^{1}_{\alpha}\rangle \otimes |\psi^{2}_{\beta}\rangle \otimes |\psi^{3*}_{\gamma}\rangle \otimes |\psi^{4*}_{\delta}\rangle
\end{equation}
and 
\begin{equation}
       A_{\alpha\beta\gamma\delta}(\tau_0) = \langle \psi^{1}_{\alpha}\psi^{2}_{\beta}|e^{-\tau_0 H_0}|\psi^{3}_{\gamma}\psi^{4}_{\delta}\rangle.
\end{equation}
As in the three-sided TFD, the expression makes sense only if a UV cutoff (lattice spacing) is introduced. In the limit of the lattice spacing going to zero, it can be shown, through a conformal map to the complex plane, that
\begin{eqnarray}
    \frac{A_{\alpha\beta\gamma\delta}(\tau_0)}{A_{IIII}(\tau_0)} = 
    2^{-\Delta_{\alpha}-\Delta_{\beta}} \left(\frac{e^{4\pi\tau_0/L_0}-1}{2e^{6\pi\tau_0/L_0}}\right)^{\Delta_{\gamma}+\Delta_{\delta}}\times \nonumber \\
    \label{eq:4_sided_As}
    \langle \phi_{\alpha}(i) \phi_{\beta}(-i) \phi_{\gamma}(ie^{-2\pi\tau_0/L_0}) \phi_{\delta}(-ie^{-2\pi\tau_0/L_0}) \rangle_{pl},
\end{eqnarray}
which involves a four-point correlation function of the CFT on the plane. As is well known in CFT, the four-point correlation function can be decomposed into different fusion channels. The fusion channels clearly agree in Eq.~\eqref{eq:4_sided_As} and the geometry in Fig.~\ref{fig:geometry}. In the limit of $\tau_0\rightarrow 0$, the first point and the third point are close, which corresponds to the so-called $t$ channel of fusion. In the opposite limit where $\tau_0\rightarrow\infty$, the last two points are close, which corresponds to the so-called $s$ channel of fusion.

Finally, the four-sided TFD state encodes the scaling dimensions and OPE coefficients in the following way,
\begin{eqnarray}
\label{eq:4_sided_As_tilde}
    &~&\frac{\tilde{A}_{\alpha\beta\gamma\delta}(\tau_1,\tau_2,\tau_3,\tau_4;\tau_0)}{{\tilde{A}}_{IIII}(\tau_1,\tau_2,\tau_3,\tau_4;\tau_0)}  \nonumber \\
    &=&e^{-\frac{2\pi\Delta_{\alpha}\tau_1}{L_1}-\frac{2\pi\Delta_{\beta}\tau_2}{L_2}-\frac{2\pi\Delta_{\gamma}\tau_3}{L_3}-\frac{2\pi\Delta_{\delta}\tau_4}{L_4}}A_{\alpha\beta\gamma\delta}(\tau_0),
\end{eqnarray}
where
\begin{eqnarray}
    &~&\tilde{A}_{\alpha\beta\gamma\delta}(\tau_1,\tau_2,\tau_3,\tau_4;\tau_0) \nonumber \\
    &=& \langle \psi^{1}_{\alpha} \psi^{2}_{\beta} \psi^{3*}_{\gamma} \psi^{4*}_{\delta}|\psi(\tau_1,\tau_2,\tau_3,\tau_4;\tau_0)\rangle
\end{eqnarray}
is the wavefunction.
\section{Construction of multi-boundary wormhole states on the lattice}
\label{sec:lat}
In the following we will construct the state $|\psi\rangle$ on the lattice and compute several entanglement quantities. In general, there are finite-size corrections if we try to compute any CFT quantity using a critical lattice model with a finite number of sites. Additional finite-size correction is present here because of the singularity of the pants geometry. We will briefly discuss how to use a "smoother" to reduce the finite-size corrections.
\subsection{Setup}
We will consider a critical quantum spin chain with Hamiltonian density $h_{i}$. The Hamiltonian on $N$ sites is
\begin{equation}
    H = \sum_{i=1}^{N} h_{i}.
\end{equation}
We will use periodic boundary conditions throughout the paper. The low-energy eigenstates are in one to one correspondence with the CFT states. The energies correspond to the scaling dimensions of the CFT \cite{blote_1986},
\begin{eqnarray}
\label{eq:Edelta}
E_{\alpha} &=& \frac{2\pi}{N} \left(\Delta_{\alpha}-\frac{c}{12}\right) + O(N^{-x}),
\end{eqnarray}
where $O(N^{-x})$ with $x>1$ denotes the finite-size correction. Note that Eq.~\eqref{eq:Edelta} may require a particular normalization of the Hamiltonian density $h_{i}$. The normalization is such that the speed of light is one. 

Next, we construct the three-sided wormhole state. Let the number of sites of the $i$-th side be $N_i$ and $N_1=N_2+N_3$. The Hilbert spaces on the three sides satisfy $\mathcal{H}_1 = \mathcal{H}_2 \otimes \mathcal{H}_3$. The spins are labelled by $i=1,2\cdots 2N_1$, where the first side contains spins $1$ to $N_1$, the second side contains spins $N_1+1$ to $N_1+N_2$ and the third side contains the rest of the spins. The Hamiltonians on the three sides are given by
\begin{eqnarray}
\label{eq:H1}
   H_1 &=& \sum_{i=1}^{N_1} h_i \\
   H_2 &=& \sum_{i=N_1+1}^{N_1+N_2} h_i \\
   \label{eq:H3}
   H_3 &=& \sum_{i=N_1+N_2+1}^{2N_1} h_i.
\end{eqnarray}
Note that periodic boundary conditions are assumed for each of the subsystem, e.g. in $H_1$ the spin 1 is identified with spin $N_1+1$, and in $H_2$ the spin $N_1+1$ is identified with the spin $N_1+N_2+1$, see Fig. \ref{fig:smoother} for an illustration.

The maximal entangled state is given by 
\begin{equation}
    |\Omega_{1,23}\rangle = \prod_{i=1}^{N_1} |\Psi_{i,N_1+i}\rangle
\end{equation}
where
\begin{equation}
    |\Psi_{i,j}\rangle =\frac{1}{\sqrt{d}}\sum_{m=1}^{d}|m_im_j\rangle
\end{equation}
is a maximal entangled state between two sites and $d$ is the dimension of the Hilbert space on each lattice site (e.g. $d=2$ for a spin-$1/2$ chain discussed below). The three-sided wormhole state can then be constructed by using the imaginary time evolution,
\begin{equation}
    |\psi_3(\tau_1,\tau_2,\tau_3)\rangle = e^{-\tau_1 H_1-\tau_2 H^{*}_2-\tau_3 H^{*}_3}|\Omega_{1,23}\rangle.
\end{equation}
As we will numerically show, this state is the lattice analogue of the three-sided wormhole state in the sense that the wavefunction
\begin{equation}
\label{eq:3_sided_As_lat}
    \tilde{A}_{\alpha\beta\gamma}(\tau_1,\tau_2,\tau_3) = \langle \psi_{\alpha} \psi^{*}_{\beta} \psi^{*}_{\gamma}|\psi_{3}(\tau_1,\tau_2,\tau_3)\rangle
\end{equation}
satisfies Eq.~\eqref{eq:3_sided_As_tilde} with finite-size corrections that decay with $N$. 

Analogously, we may construct the four-sided wormhole state. Let the number of sites of the four sides be $N_1,N_2,N_3$ and $N_4$, and $N_1+N_2=N_3+N_4=N_0$. The Hilbert spaces satisfy  $\mathcal{H}_0 = \mathcal{H}_1 \otimes \mathcal{H}_2 = \mathcal{H}_3 \otimes \mathcal{H}_4$. Then the Hamiltonian on the circles are
\begin{eqnarray}
   H_1 &=& \sum_{i=1}^{N_1} h_i \\
   H_2 &=& \sum_{i=N_1+1}^{N_0} h_i \\
   H_3 &=& \sum_{i=N_0+1}^{N_0+N_3} h_i \\
   H_4 &=& \sum_{i=N_0+N_3+1}^{2N_0} h_i \\
   H_0 &=& \sum_{i=1}^{N_0} h_i.
\end{eqnarray}
One starts with the maximal entangled state 
\begin{equation}
    |\Omega_{12,34}(0)\rangle = \prod_{i=1}^{N_0} |\Psi_{i,N_0+i}\rangle 
\end{equation}
on $\mathcal{H}_{0} \otimes \mathcal{H}^{*}_{0}$, then a TFD state can be constructed by imaginary time evolution with $H_0$,
\begin{equation}
    |\Omega_{12,34}(\tau_0)\rangle = e^{-\tau_0 H_0/2}e^{-\tau_0 H^{*}_0/2} |\Omega_{12,34}(0)\rangle
\end{equation}
The four-sided wormhole state can be obtained by further imaginary time evolution on four subsystems,
\begin{eqnarray}
        &~&|\psi_4(\tau_1,\tau_2,\tau_3,\tau_4;\tau_0)\rangle \nonumber \\
        &=& e^{-\tau_1 H_1-\tau_2 H_2-\tau_3 H^{*}_3-\tau_4 H^{*}_4}|\Omega_{12,34}(\tau_0)\rangle.
\end{eqnarray}
Again, as we will show numerically, the lattice state is analogous to the continuous counterpart which satisfies Eq.~\eqref{eq:4_sided_As_tilde} for primary states.

\subsection{Finite-size corrections and the smoother}
In this subsection we analyze the finite-size corrections in Eq.~\eqref{eq:3_sided_As_lat}. The imaginary time evolution contributes a factor $e^{-\tau_1 E_\alpha-\tau_2 E_\beta -\tau_3 E_{\gamma}}$ to the wavefunction, since the states $|\psi_{\alpha}\rangle, |\psi^{*}_{\beta}\rangle$ and $|\psi^{*}_{\gamma}\rangle$ are eigenstates of $H_1$, $H_2$ and $H^{*}_3$, respectively,
\begin{equation}
    \tilde{A}_{\alpha\beta\gamma} = e^{-\tau_1 E_\alpha-\tau_2 E_\beta -\tau_3 E_{\gamma}} A_{\alpha\beta\gamma}
\end{equation}
where
\begin{equation}
    A_{\alpha\beta\gamma} = \langle \psi^{1}_{\alpha} \psi^{2*}_{\beta} \psi^{3*}_{\gamma}|\Omega_{1,23}\rangle.
\end{equation}
Identifying the circumferences $L$'s with the number of spins $N$'s, the factor agrees with the prefactor in Eq.~\eqref{eq:3_sided_As_tilde} up to the finite-size correction in the energies, Eq.~\eqref{eq:Edelta}. 

Additional finite-size corrections come from $A_{\alpha\beta\gamma}$. It turns out that the finite-size corrections in $A_{\alpha\beta\gamma}$ are much stronger than the finite-size corrections in the energies. We will make use of \textit{smoothers} to reduce the finite-size corrections in $A_{\alpha\beta\gamma}$, analogous to those of Refs.~\cite{Milsted:2018san,Milsted:2018vop,Milsted:2018yur}. 

\begin{figure}
    \centering
    \includegraphics[width=0.83\linewidth]{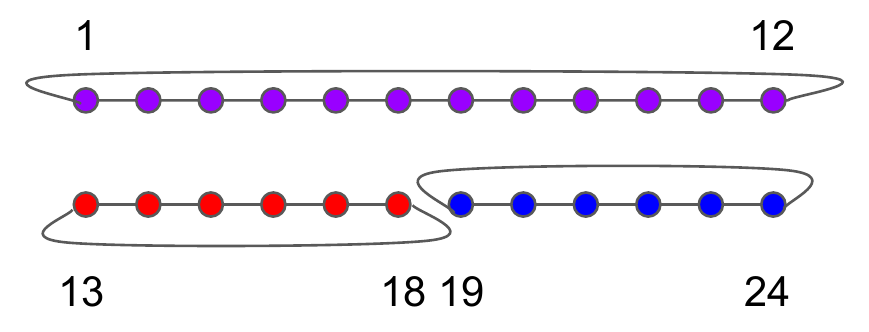}
    \includegraphics[width=0.8\linewidth]{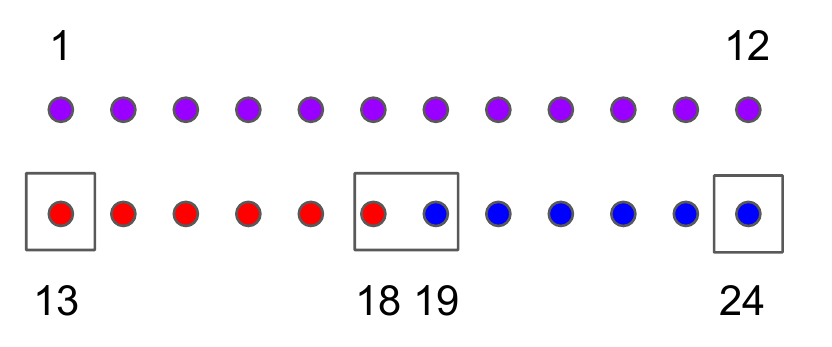}
    \caption{Top: An illustration of the spin configuration in the three-sided TFD state. The purple circles represent spins in subsystem 1. The reds represent subsystem 2 and the blues represent subsystem 3. The lines represent nearest-neighboring sites in the Hamiltonians Eqs.~\eqref{eq:H1}-\eqref{eq:H3}. Bottom: An illustration of the region of spins where the smoother acts. The sizes are $N_1=12, N_2=6,N_3=6$ and the smoother size is $l=1$.}
    \label{fig:smoother}
\end{figure}

In order to obtain the smoother, first note that
\begin{equation}
\label{eq:3_sided_As_lat_def}
    A_{\alpha\beta\gamma} = \langle \psi^{1}_{\alpha}|\psi^{2}_{\beta}\psi^{3}_{\gamma}\rangle
\end{equation}
by definition of the maximal entangled state. The strong finite-size correction is due to the sharp change of the boundary condition when we take the overlap. Let $U_{23}$ be a local unitary operator of size $4l$ which acts on spins on $\mathcal{H}_2$ and $\mathcal{H}_3$ near the boundary condition change, i.e., the spins labelled by $N_1+1$ to $N_1+l$, $N_1+N_2-l+1$ to $N_1+N_2+l$ and $2N_1-l+1$ to $2N_1$. See Fig.~\ref{fig:smoother} for an illustration. We find $U_{23}$ by an optimization
\begin{equation}
    \max_U |\langle I^{1} |U_{23}| I^{2} I^{3}\rangle|^2.
\end{equation}
The physical meaning of the unitary $U_{23}$ is that it "smoothens" the connection of the copy $2$ and copy $3$ such that the resulting state overlaps most with the low-energy subspace of the copy $1$. It is important to note that the unitary $U$ is optimized only once with a fairly large system and the same unitary is applied for all system sizes. We will show numerically that the wavefunction $\tilde{A}_{\alpha\beta\gamma}$ satisfies Eq.~\eqref{eq:3_sided_As_tilde} with a smaller finite-size correction. 
\subsection{The Ising model}
The critical Ising model is defined on a spin-1/2 lattice,
\begin{equation}
    H = -\frac{1}{2}\left(\sum_{i=1}^{N} X_{i} X_{i+1} + \sum_{i=1}^{N} Z_i\right),
\end{equation}
where periodic boundary conditions are assumed. The normalization constant $1/2$ is such that Eq.~\eqref{eq:Edelta} holds.

Using a Jordan-Wigner transformation one may map the spin model to the free fermion model on $2N$ Majorana fermions $\gamma_i$,
\begin{equation}
    \{\gamma_{i},\gamma_{j}\}=2\delta_{ij}.
\end{equation}
The Hamiltonian is 
\begin{equation}
\label{eq:H_Ising}
    H = -\frac{i}{2}\sum_{i=1}^{2N} \gamma_i \gamma_{i+1}.
\end{equation}
There are two possible boundary conditions, the NS boundary condition $\gamma_{2N+1}=-\gamma_{1}$ or R boundary condition $\gamma_{2N+1}=\gamma_{1}$. The eigenstates of the spin Hamiltonian with even/odd $\mathbb{Z}_2$ symmetry are mapped to eigenstates of the fermion model with NS/R boundary condition. 

In the following we will use the free fermion model instead of the spin model to simplify numerical simulations. It should be noted that a numerical simulation of the spin model is also possible, by using tensor network techniques developed in Refs.~\cite{Zou_2017,Milsted:2018vop}. 

\section{Free fermion techniques}
\label{sec:tech}
In this section we outline the free fermion techniques that are used to construct the multi-boundary wormhole state for the free fermion model. We first review the correlation matrix formalism and then introduce two new techniques, the imaginary time evolution and the construction of the TFD state.  
\subsection{Review of the correlation matrix formalism}
Here we work with Gaussian Majorana fermion states. A Gaussian state (either pure or mixed) is completely determined by the two-point correlation functions,
\begin{equation}
    M_{ij} = \Tr(-i\rho (\gamma_i \gamma_j-\delta_{ij})) 
\end{equation}
The matrix $M$ is known as the correlation matrix. It is a real, skew-symmetric, $2N\times 2N$ matrix. It can be block-diagonalized by an orthogonal matrix,
\begin{equation}
\label{eq:Schur}
    M = O\Gamma O^{T},
\end{equation}
where $O\in O(2N)$ and
\begin{equation}
\label{eq:Schur_diag}
    \Gamma = \oplus_{i=1}^N n_i (i\sigma^y) \equiv \diag(n_i) \otimes i\sigma^y
\end{equation}
is block diagonal and $n_i \in [-1,1]$. A pure state has $n_i=\pm 1$ and thus $M^2=-I$. The Von-Neumann entropy of the density matrix is given by \cite{Peschel_2009}
\begin{equation}
    S(\rho) \equiv -\Tr(\rho\log\rho) = H\left(\frac{1+n_i}{2}\right)+H\left(\frac{1-n_i}{2}\right),
\end{equation}
where 
\begin{equation}
    H(x)=-\sum_i x_i \log x_i.
\end{equation}
The expectation value of the fermion parity 
\begin{equation}
    P = (-i)^N \prod_{i=1}^N \gamma_{2i-1}\gamma_{2i}
\end{equation}
is given by 
\begin{equation}
    \Tr(\rho P) = \Pf(M)
\end{equation}
as a result of the Wick theorem, where $\Pf$ denotes the Pfaffian.

Given two states $\rho_1,\rho_2$ with correlation matrices $M_1,M_2$, if at least one state is pure, then their overlap is \cite{Bravyi_2005,Bravyi_2017}
\begin{equation}
\label{eq:fermion_ov}
    \Tr(\rho_1\rho_2) = 2^{-N}|\Pf{(M_1+M_2)}|.
\end{equation}
This expression will be useful for computing the wavefunctions from the correlation matrix.

Let $H$ be a quadratic Hamiltonian
\begin{equation}
    H = -\frac{i}{4}\sum_{ij} h_{ij} \gamma_i \gamma_j,
\end{equation}
where $h$ is a real, skew-symmetric, $2N\times 2N$ matrix. The matrix $h$ admits a Schur decomposition
\begin{equation}
    h = O (\diag(\epsilon_i)\otimes i\sigma^y) O^{T}.
\end{equation}
then the thermal state $\rho_{th}(\tau) \sim e^{-\tau H}$ has correlation matrix
\begin{equation}
    M_{th}(\tau) = O (\diag(-tanh{(\tau\epsilon_i/2)})\otimes i\sigma^y) O^{T}
\end{equation}
In the limit of $\tau\rightarrow +\infty$, we obtain the correlation matrix for the ground state
\begin{equation}
    M_{gs} = O (\diag(-\mathrm{sgn}{(\epsilon_i)})\otimes i\sigma^y) O^{T}.
\end{equation}
Every eigenstate of the Hamiltonian has a correlation matrix of the form
\begin{equation}
\label{eq:fermion_eVs}
    M(\{n_i\}) = O (\diag(n_i)\otimes i\sigma^y) O^{T},
\end{equation}
where each eigenstate corresponds to a choice of $n_i \in \{-1,1\}$ for $i$ from $1$ to $N$. The energy of that eigenstate is given by
\begin{equation}
    E(\{n_i\}) = \sum_{i=1}^N n_i \epsilon_i/2.
\end{equation}
The ground state corresponds to $n_i = -\mathrm{sgn}(\epsilon_i)$.

Finally, if we have two Gaussian fermion states $|\psi_1\rangle$ and $|\psi_2\rangle$ with correlation matrices $M_1$ and $M_2$, then the state of composite system has the correlation matrix $M_1 \oplus M_2$. 

\subsection{Imaginary time evolution}
Let $H$ be a quadratic Hamiltonian and $|\psi\rangle$ be a pure state with correlation matrix $M$. Consider the imaginary time evolution
\begin{equation}
    |\psi(\tau)\rangle = \frac{1}{\sqrt{Z(\tau)}}e^{-\tau H}|\psi\rangle
\end{equation}
where $Z(\tau)$ is a normalization constant. The resulting state $|\psi(\tau)\rangle$ is a Gaussian state with the correlation matrix
\begin{eqnarray}
\label{eq:evol_imag}
    M(\tau) = M_{th}(2\tau) - G(\tau) (M+M_{th}(2\tau))^{-1} G(\tau),
\end{eqnarray}
where
\begin{equation}
    G(\tau) = \sqrt{I+M^2_{th}(2\tau)}.
\end{equation}
Note that at $\tau=0$, $M(0) = -M^{-1} = M$, as $M^2=-I$ for a pure state.  As $\tau\rightarrow \infty$, the thermal state goes into the ground state, $M_{th}(\tau)\rightarrow M_{gs}$ and $M(\tau)\rightarrow M_{gs}$.

The normalization constant is 
\begin{equation}
    Z(\tau) = \prod_{i} \cosh(\tau\epsilon_i) |\Pf(M+M_{th}(2\tau))|.
\end{equation}

\subsection{Construction of TFD state}
Let $M$ be the correlation matrix of a density matrix $\rho$ on $2N$ Majorana fermions, and the spectral decomposition of the density matrix be
\begin{equation}
    \rho = \sum_{\alpha} \lambda^2_{\alpha} |\alpha\rangle\langle \alpha|. 
\end{equation}
The canonical purification is a pure state on $4N$ Majorana fermions defined by
\begin{equation}
    |\sqrt{\rho}\rangle \equiv \sum_{\alpha} \lambda_{\alpha} |\alpha\rangle \otimes |\alpha^{*}\rangle.
\end{equation}
Note that if $\rho$ is the thermal state of a local Hamiltonian, then the canonical purification coincides with the TFD state. As we show in the appendix, the correlation matrix of the canonical purification is
\begin{equation}
\label{eq:can_pur}
    M_{\sqrt{\rho}} = 
    \begin{bmatrix}
    M & \tilde{M} \\
    -\tilde{M} & -M 
    \end{bmatrix}
\end{equation}
where 
\begin{equation}
    \tilde{M} = O \left(\diag\left({\sqrt{1-n^2_i}}\right) \otimes I_2 \right) O^T,
\end{equation}
given the $n_i$ and the matrix $O$ in Eqs.~\eqref{eq:Schur} and \eqref{eq:Schur_diag}. In particular, if the original state $\rho$ is maximally mixed state with $M=0$, then the canonical purification is the maximal entangled state $|\Omega\rangle$. The correlation matrix of the state $|\Omega\rangle$ is
\begin{equation}
\label{eq:corr_Omega}
    M_{\Omega} =     \begin{bmatrix}
    0_{2N} & I_{2N} \\
    -I_{2N} & 0_{2N}
    \end{bmatrix}
\end{equation}
In order to compute the correlation matrix for the multi-boundary TFD state, we need to use the imaginary time evolution Eq.~\eqref{eq:evol_imag} and the correlation matrix Eq.~\eqref{eq:corr_Omega} for the initial state. The generic expression Eq.~\eqref{eq:can_pur} is also useful when we compute the multi-partite entanglement measure next section.

\section{Numerical results}
\label{sec:num}
In this section we show the numerical results for the three- and four-sided TFD state for the Ising (free fermion) model. We first compute the wavefunctions and show that they agree with the CFT correlation functions as in Eq.~\eqref{eq:3_sided_As} and Eq.~\eqref{eq:4_sided_As}. We then compute entanglement quantities in the state and show their dependence on the parameters. We find ranges of the parameters where tripartite entanglement is significant or negligible.
\subsection{CFT correlation functions}
The Ising CFT has three primaries $I,\sigma,\varepsilon$ with scaling dimensions $\Delta_{I} = 0, \Delta_{\sigma} =1/8$ and $\Delta_{\varepsilon} =1$. The OPE coefficients are completely symmetric in their three indices and the nonzero OPE coefficients are $C_{I\alpha\beta}=\delta_{\alpha\beta}$ and $C_{\sigma\sigma\varepsilon}=1/2$ \cite{Belavin_1984,Friedan_1984}. The value of the RHS of Eq.~\eqref{eq:3_sided_As} is shown in Table \ref{tab:3_sided_As}. 

Numerically, the overlaps $A_{\alpha\beta\gamma}$ can be computed in the following way. First, we compute the correlation matrix of the eigenstates of the free fermion model with $2N$ fermions and $4N$ fermions using Eq.~\eqref{eq:fermion_eVs}. The ground state of the Hamiltonian Eq.~\eqref{eq:H_Ising} with the NS boundary condition is identified with $|I\rangle$ and the first excited state with even fermion parity is identified with $|\varepsilon\rangle$. Then, we compute the correlation matrix of the ground state of the Hamiltonian Eq.~\eqref{eq:H_Ising} with the R boundary condition. The ground state is two-fold degenerate, and the one with odd fermion parity is identified with $|\sigma\rangle$. Next, we optimize the smoother and apply it to the states. Finally, we compute the overlap Eq.~\eqref{eq:3_sided_As_lat_def} using Eq.~\eqref{eq:fermion_ov}. 

\begin{figure}
    \centering
    \includegraphics[width=0.98\linewidth]{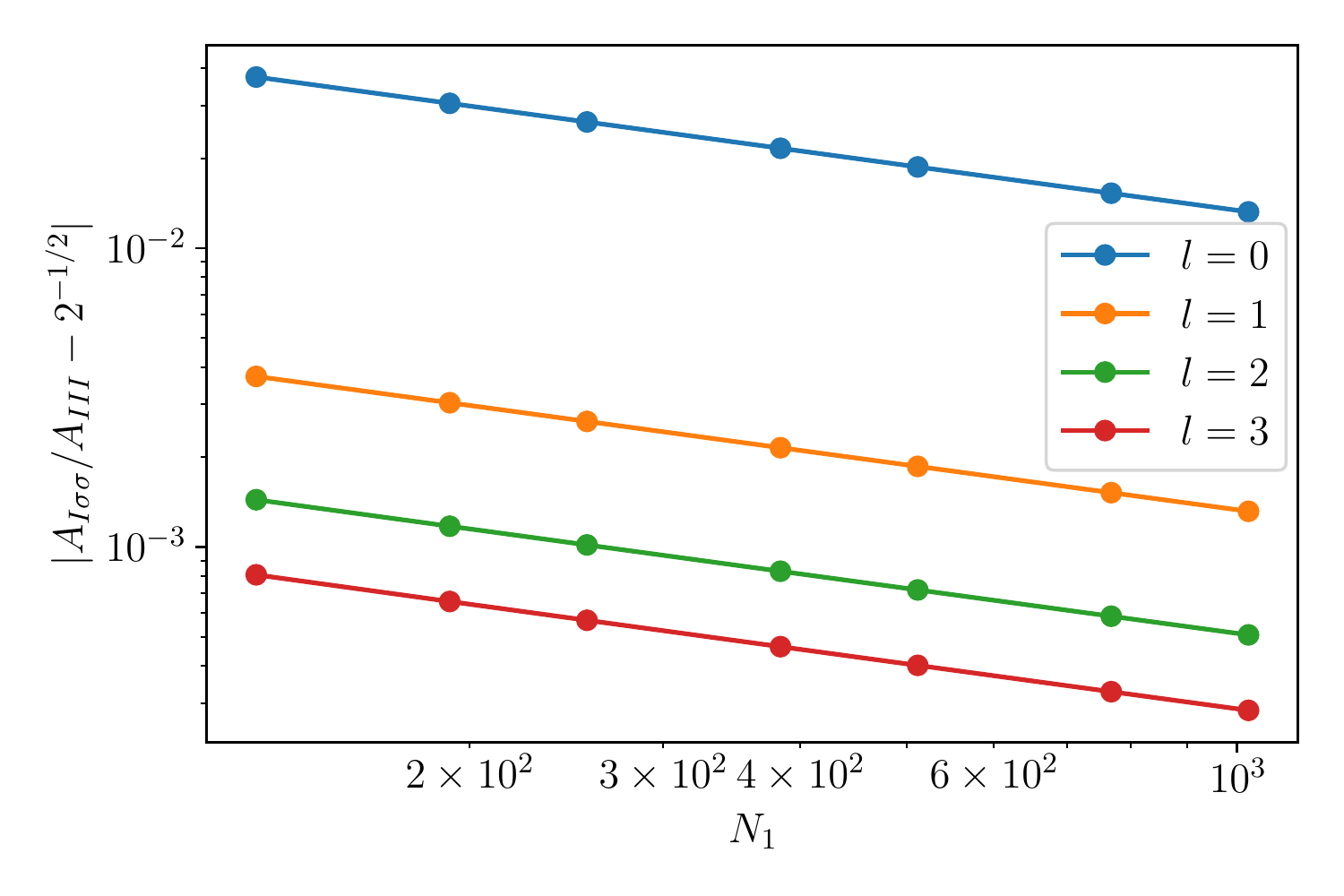}
    \includegraphics[width=0.98\linewidth]{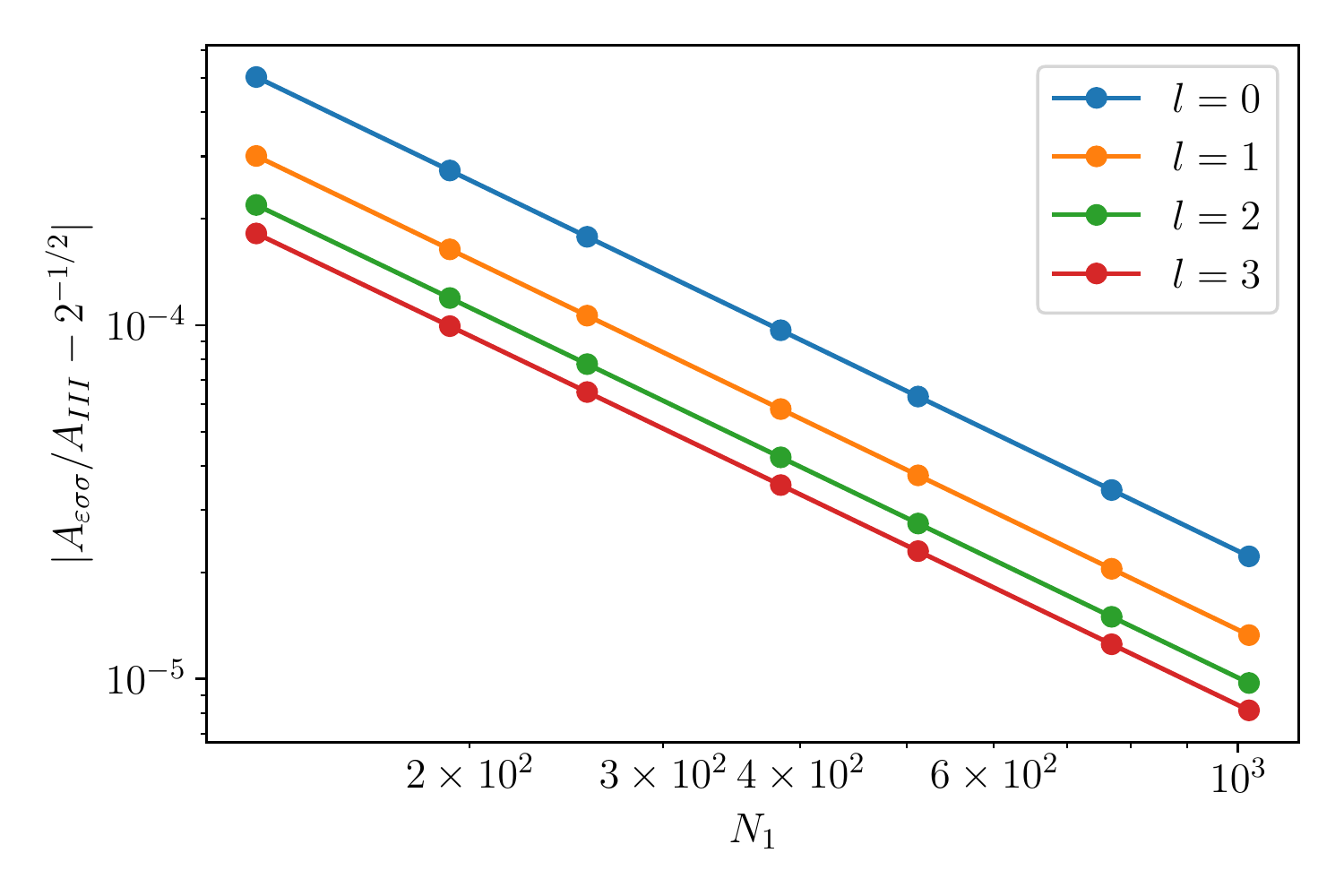}
    \caption{Finite-size corrections of the wavefunction $A_{\alpha\beta\gamma}$ of the three-sided TFD state with different sizes of smoothers. We see that the smoothers reduce the finite-size corrections significantly but leave the scaling with the system sizes unchanged.} 
    \label{fig:3sided_As}
\end{figure}
The result is shown in Tab.~\ref{tab:3_sided_As} and Fig.~\ref{fig:3sided_As}. We see that all OPE coefficients are reproduced accurately. In addition, we see that the smoothers reduce the finite-size corrections significantly. More precisely, although the smoothers do not change the scaling of the finite-size corrections with the system sizes but they reduce the prefactors significantly. With the smoother size $l=2$, the prefactors in front of the finite-size corrections are reduced by two orders of magnitude. 

\begin{table}[]
    \centering
    \begin{tabular}{|c|c|c|c|}
    \hline
       $\psi_{\beta},\psi_{\gamma}$ & $\psi_{\alpha}$ & $2^{-2\Delta_{\beta}-2\Delta_\gamma+\Delta_{\alpha}}C_{\alpha\beta\gamma}$ (CFT) & $A_{\alpha\beta\gamma}/A_{III}$ (Lattice)  \\ \hline
        $\sigma,\sigma$ & $I$ & $2^{-1/2} \approx 0.7071$ & 0.7065  \\ \hline
        $\sigma,I$ & $\sigma$ & $2^{-1/8} \approx 0.9167 $& 0.9170 \\ \hline
        $\sigma,\varepsilon$ & $\sigma$ & $2^{-25/8}\approx 0.1139 $& 0.1146  \\ \hline
        $\varepsilon,I$ & $\varepsilon$ & $1/2=0.5$& 0.5000 \\ \hline
        $\varepsilon,\varepsilon$ & $I$ & $1/16=0.0625$ & 0.0618 \\ \hline
    \end{tabular}
    \caption{The wavefunctions of the three-sided wormhole for the Ising model, CFT versus lattice results. The lattice result is computed with $N_1=2N_2=2N_3=512$ and the smoother size $l=2$.}
    \label{tab:3_sided_As}
\end{table}
The four-point correlation function of the Ising CFT is known exactly \cite{Belavin_1984}. For simplicity, we consider the correlation function of four $\sigma$ fields. As shown in appendix, the RHS of Eq.~\eqref{eq:4_sided_As} equals 
\begin{equation}
\label{eq:4_sided_As_CFT}
    \frac{A_{\sigma\sigma\sigma\sigma}(\tau_0)}{A_{IIII}(\tau_0)} = \frac{1}{2}\sqrt{\frac{1-e^{-4\pi\tau_0/N_0}}{tanh (\pi \tau_0/N_0)}}
\end{equation}

\begin{figure}
    \centering
    \includegraphics[width=0.98\linewidth]{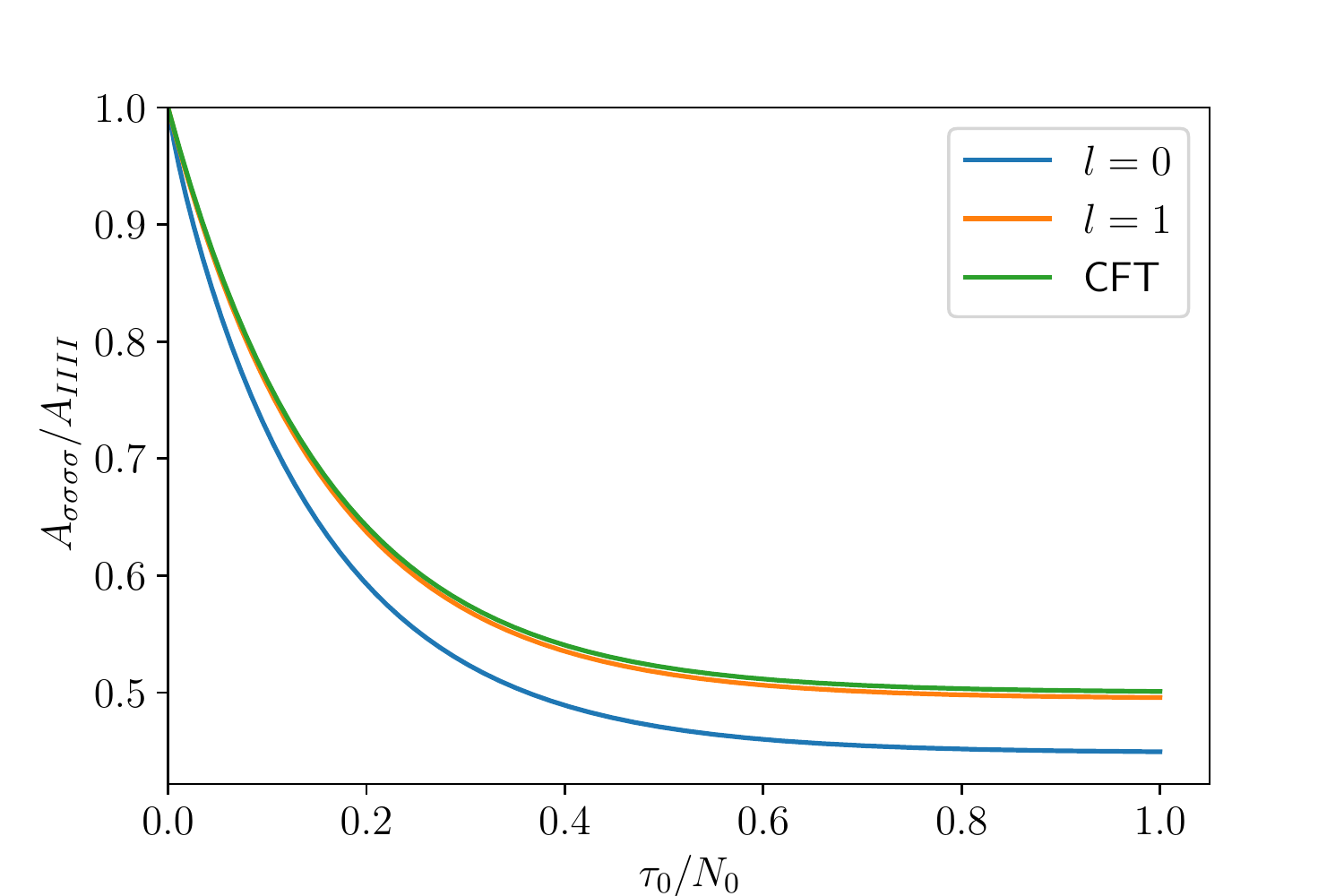}
    \caption{The ratio between wavefunctions $A_{\sigma\sigma\sigma\sigma}/A_{IIII}$ for the free fermion model. The lattice computation is done with $N_0=2N_1=2N_2=2N_3=2N_4=128$ and smoother sizes $l=0$ (blue) and $l=1$ (orange). The CFT expression Eq.~\eqref{eq:4_sided_As_CFT} is shown in the green curve. }
    \label{fig:4_sided_corrs}
\end{figure}

Numerically, the wavefunction $A_{\alpha\beta\gamma\delta}$ for the free fermion model in the following way. First we compute the correlation matrices of the four states in the overlap. Then we construct the correlation matrix of the TFD state $|\Omega_{12,34}(\tau_0)\rangle$ and apply the smoother. Finally we compute the overlap $A_{\alpha\beta\gamma\delta}(\tau_0) = \langle \psi^{1}_{\alpha} \psi^{2}_{\beta} \psi^{3*}_{\gamma}\psi^{4*}_{\delta}|\Omega_{12,34}(\tau_0)\rangle$. The result is shown in Fig.~\ref{fig:4_sided_corrs}. Similar to the OPE coefficients, the four point correlation function is reproduced accurately. Again, we see that smoothers significantly reduces the finite-size corrections. The lattice computation agrees well with the CFT expression with a small smoother size $l=1$. 
\subsection{Entanglement quantities}
In the context of holography, entanglement properties of the CFT state on the boundary reflect geometric properties of the bulk \cite{Ryu_2006,Blanco_2013,Takayanagi_2017,Dutta_2019,Nguyen_2017}. For example, the Ryu-Takayanagi formula \cite{Ryu_2006} connects the entanglement entropy of the boundary CFT to the area of the minimal surface in the bulk. However, the entanglement entropy only characterizes bipartite entanglement and multi-partite entanglement plays an important role in many-body systems. There are vast number of multi-partite entanglement quantities, and we focus on the reflected entropy $S_R$ \cite{Dutta_2019}. Given a pure state on three parties $1,2$ and $3$, we first trace out subsystem $1$ to obtain the reduced density matrix $\rho_{23}$. The reflected entropy is defined as
\begin{equation}
    S_R(2:3) = S_{22^{*}}(|\sqrt{\rho}\rangle_{232^{*}3^{*}}),
\end{equation}
where $|\sqrt{\rho}\rangle_{232^{*}3^{*}}$ is known as the canonical purification, which is the square root of the density matrix regarded as a state in the doubled Hilbert space $\mathcal{H}_{23}\otimes \mathcal{H}^{*}_{23}$. The quantity is interesting because it is conjectured to be dual to the entanglement wedge cross-section in the context of holography \cite{Dutta_2019}. Furthermore, it has been shown that the derived quantity
\begin{equation}
    h(2:3) = S_R(2:3) - I(2:3),
\end{equation}
represents the amount of multi-partite entanglement beyond the GHZ type \cite{Zou_202002}. We will use the quantity $h$ as a measure of multi-partite entanglement.

The three-sided TFD state is naturally split into three subsystems. we will study how the bipartite and multi-partite entanglement quantities change with the parameters $\tau_1,\tau_2$ and $\tau_3$. The lattice construction enables us to compute them numerically without referring to holographic conjectures. 

Before going into the numerical results, it is useful to consider several limits as in Ref.~\cite{Balasubramanian_2014} . If one of $\tau_i$'s goes to infinity, then that the copy of CFT is projected onto the ground state and there is no entanglement between this copy and the rest two copies. In the opposite limit, if $\tau_1\rightarrow 0$, then the reduced density matrix $\rho_{23}$ becomes the tensor product of two thermal state $\rho_{23}\sim e^{-2\tau_2 H_2}e^{-2\tau_3 H_3}$ and $S_R(2:3)=I(2:3)=0$. It is therefore expected that multi-partite entanglement is only significant when $\tau_1$ is not too large or too small.

\begin{figure}
    \centering
    \includegraphics[width=\linewidth]{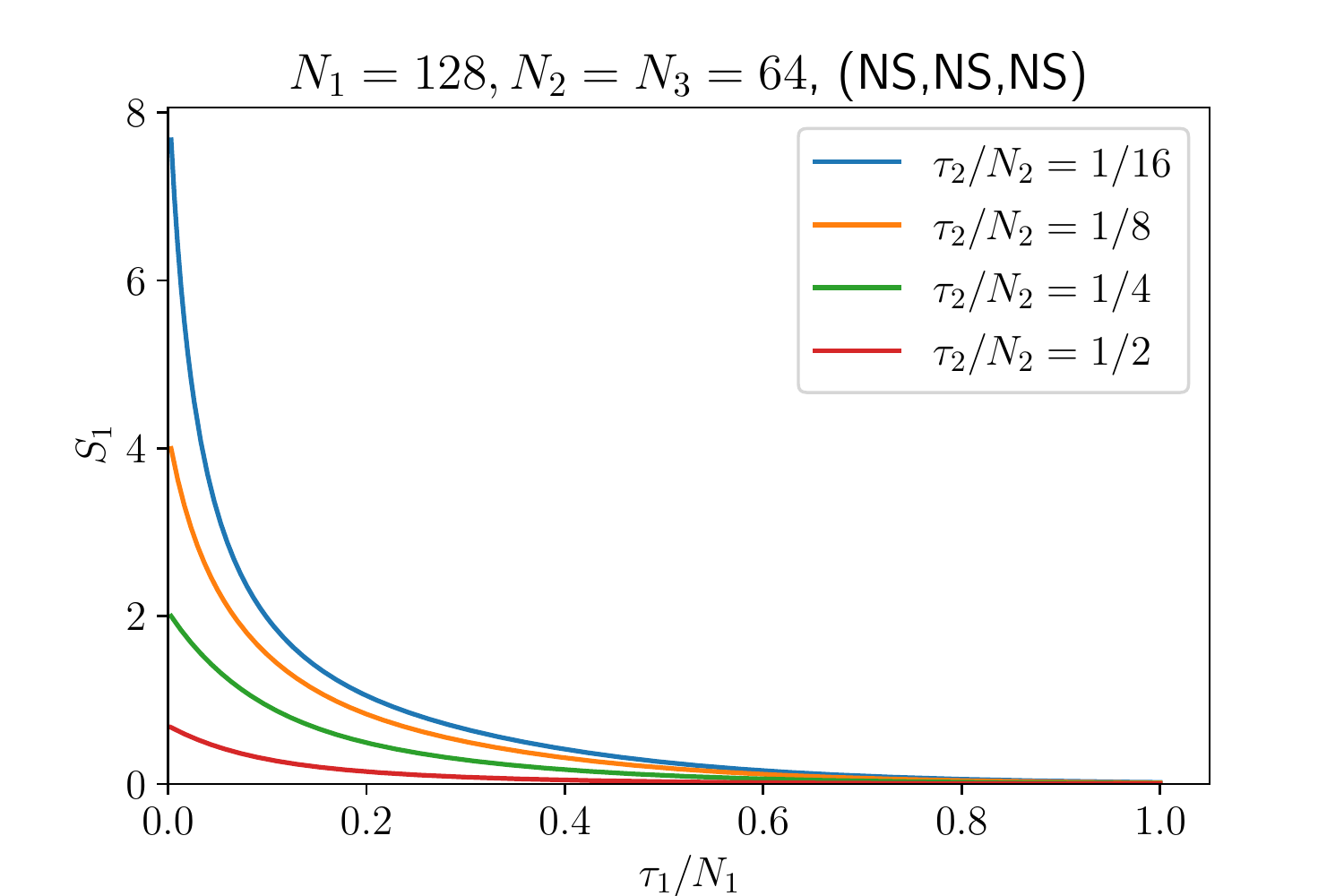}
    \includegraphics[width=\linewidth]{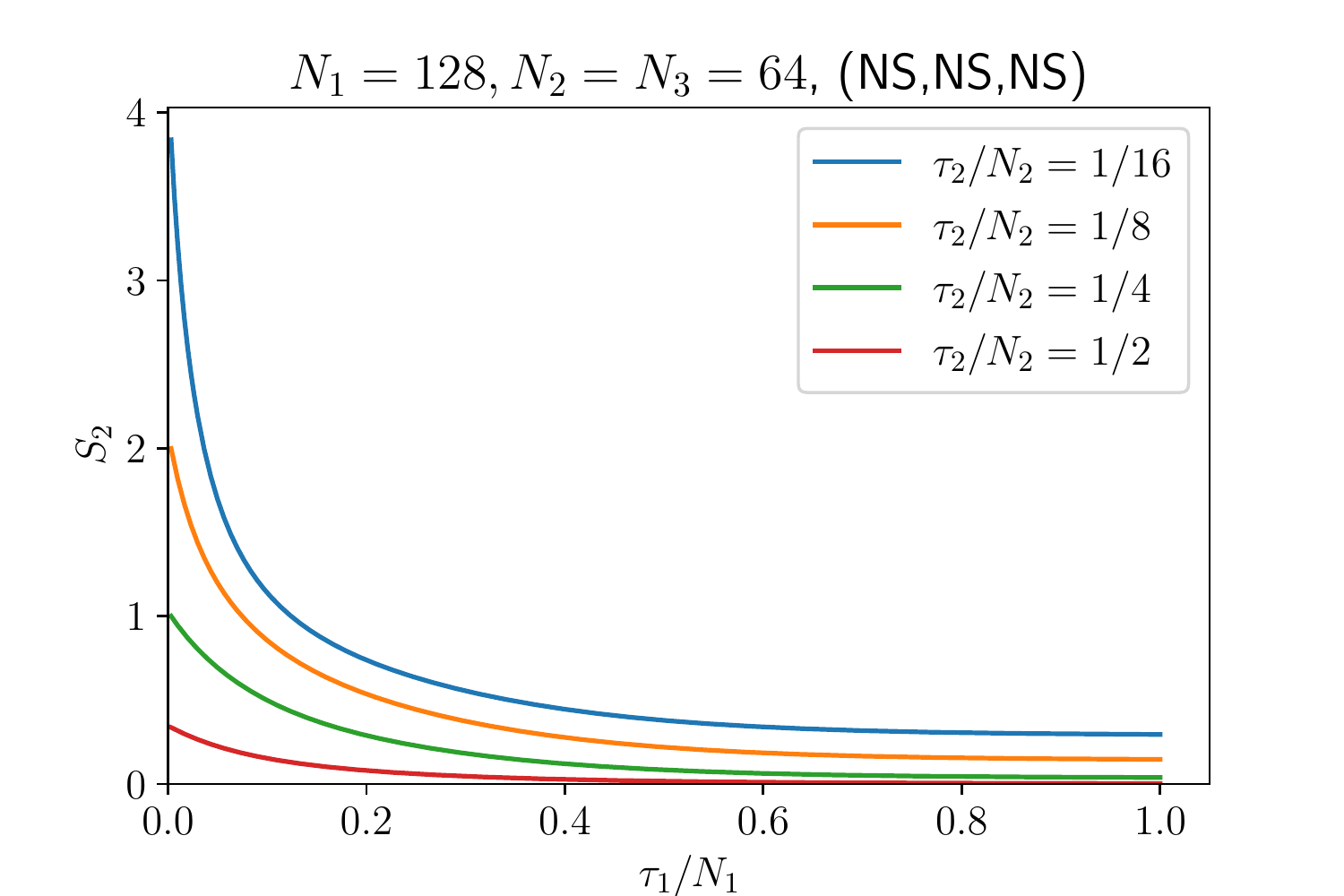}
    \caption{The entanglement entropy $S_1$ and $S_2$ for the three-sided TFD state.}
    \label{fig:EEs}
\end{figure}

\begin{figure}
    \centering
    \includegraphics[width=\linewidth]{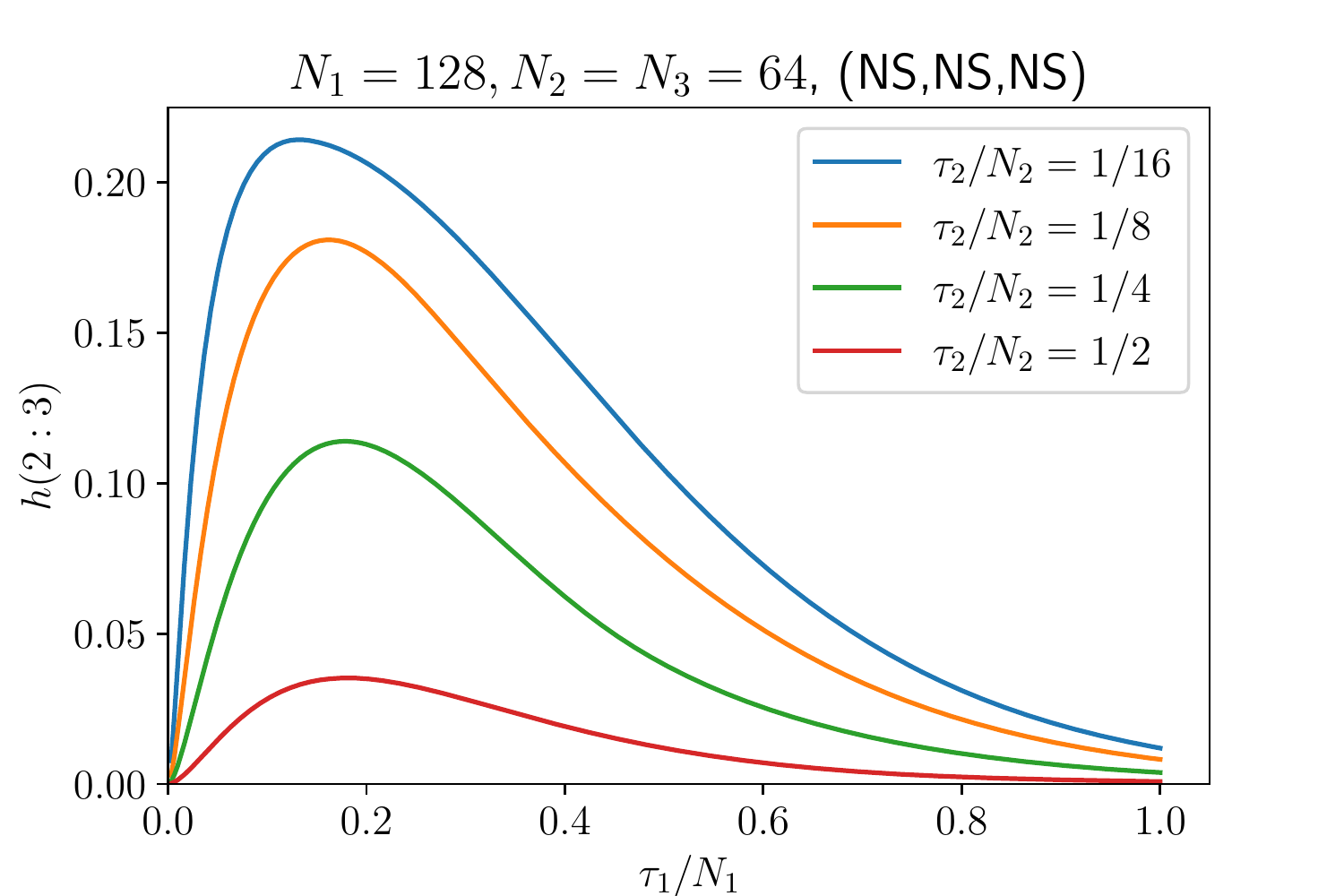}
    \caption{The tripartite entanglement measure $h$ for the three-sided TFD state.}
    \label{fig:h}
\end{figure}

For simplicity we consider the case where $\tau_2=\tau_3$ and we fix the boundary conditions of the three sides to be all NS. The entanglement quantities with respect to $\tau_1$ and $\tau_2$ is shown in Fig.~\ref{fig:EEs} and Fig.~\ref{fig:h}.

The bipartite entanglement entropy $S_1$ and $S_2$ are monotonically decreasing functions of $\tau_1$ and $\tau_2$. As $\tau_1\rightarrow\infty$, $S_1$ goes to zero as the reduced density matrix $\rho_1$ is projected onto the ground state, and $S_2$ goes to a finite value that decreases with $\tau_2$. The behaviors are in qualitative agreement with holographic entanglement entropy \cite{Balasubramanian_2014}. As for the tripartite entanglement measure $h(2:3)$, we observe that it peaks roughly at $\tau_1/N_1\approx 0.2$ for fixed $\tau_2$ and the maximum decreases with $\tau_2$. In a holographic CFT, such a peak still exists, but the peak is sharp \cite{Rath} rather than smooth for low-$c$ CFTs such as the Ising CFT.
\section{Conclusion}
\label{sec:concl}
In this paper we have introduced a class of multi-boundary generalizations of TFD states. We have found how multi-point correlation functions of the CFT are encoded in the wavefunctions. Furthermore, we have proposed lattice constructions of the multi-boundary TFD state for critical quantum spin chains, and introduced smoothers that significantly reduce the finite-size corrections. Our lattice construction allows, on the one hand, for numerical explorations of the corresponding wavefunctions. They also lead to a scheme to numerically extract conformal data such as operator product expansion coefficients OPE of a CFT by computing wavefunction overlaps (as opposed to related techniques in Refs. \cite{Zou_2019,Zou_2019_02}, which require computing a lattice version of the CFT primary operators). Finally, for illustrative purposes, we have numerically constructed the states for the Ising CFT using novel free fermion techniques and computed entanglement quantities.

The multi-boundary TFD states defined in this work are associated with path integral on surfaces that are flat except in a neighborhood around the singular points. This is mostly due to simplification in numerical techniques. More sophisticated tensor network techniques, such as a imaginary time evolution combined with MERA \cite{Milsted:2018san,Milsted:2018yur,Milsted:2018vop} can be used to obtain a lattice approximation to path integrals on more general, curved manifolds. In addition to correlation functions on the plane, one can also compute correlation functions on more complicated manifolds by combining several pants-like geometries. For example, the reduced density matrix of the first subsystem in the three-sided TFD state encodes two-point correlation functions on the torus. 

Finally, the states considered in this paper may have holographic interpretations in the context of holographic CFTs. Our work enables a direct computation of entanglement quantities, which may be compared with holography qualitatively or even quantitatively. 

\acknowledgements

We thank Sergey Bravyi and Qi Hu for useful discussions. We are particularly grateful to Pratik Rath for discussions about the multi-boundary wormhole state in holography. 
Sandbox is a team within the Alphabet family of companies, which includes Google, Verily, Waymo, X, and others.
GV is a CIFAR fellow in the Quantum Information Science Program and a Distinguish Visiting Research Fellow at Perimeter Institute. Research at Perimeter Institute is supported by
the Government of Canada through the Department of Innovation, Science and Economic Development
and by the Province of Ontario through the Ministry of Research, Innovation and Science.

\bibliography{BH3}

\appendix
\section{CFT path integrals}
In this appendix we derive the overlap formula involving three or four CFT primary states. 
\subsection{In and out states}
Let us consider a cylinder with circumsference $L$. A CFT primary state ket (bra) can be created by inserting a primary operator in the past/future infinity of the cylinder. 
\begin{eqnarray}
\label{eq:op_state_cyl}
    |\phi^{\cyl}_{\alpha}\rangle &=& \phi^{\cyl}_{\alpha}(-\infty)|I^{\cyl}\rangle. \\
    \langle\phi^{\cyl}_{\alpha}| &=&  \langle I^{\cyl}| \phi^{\cyl}_{\alpha}(+\infty),
\end{eqnarray}
where
\begin{eqnarray}
    \phi^{\cyl}_{\alpha}(\pm \infty) \equiv \left(\frac{2\pi}{L}\right)^{-\Delta_{\alpha}} \lim_{\tau\rightarrow \pm\infty} e^{\pm \frac{2\pi}{L}\Delta_{\alpha}\tau} \phi^{\cyl}_{\alpha}(\tau,0).
\end{eqnarray}
The operator-state correspondence can be expressed in terms of path integrals. The wavefunctionals on the $\tau=0$ circle are
\begin{eqnarray}
    &~&\langle \phi(x)|\phi^{cyl}_{\alpha}\rangle   \nonumber \\
    &=&  \frac{1}{\sqrt{Z^{cyl}}} \int_{\substack{\tau<0, ~cyl \\ \Phi(0,x) = \phi(x)}} D\Phi\, \phi^{\cyl}_{\alpha}(-\infty) e^{-S}. \\
    &~&\langle \phi^{cyl}_{\alpha}|\phi(x)\rangle   \nonumber \\
    &=&  \frac{1}{\sqrt{Z^{cyl}}} \int_{\substack{\tau>0, ~cyl \\ \Phi(0,x) = \phi(x)}} D\Phi\, \phi^{\cyl}_{\alpha}(+\infty) e^{-S}. 
\end{eqnarray}
Taking the overlap is equivalent to gluing the path integral. We will analyze the cases of three and four boundaries separately.
\subsection{Overlap involving three states}
Let $L$ be the circumference of circle 1 and the other two circles have circumference $L/2$ The overlap ratio can be written as a three-point correlator on the pants geometry (Fig 1),
\begin{equation}
    \frac{A_{\alpha\beta\gamma}}{A_{III}} = \langle \phi^{1}_{\alpha}(+\infty)\phi^{2}_{\beta}(-\infty)\phi^{3}_{\gamma}(-\infty)\rangle_{pants}.
\end{equation}
We use the conformal map $w=f(z)$ to map the pants geometry to the complex plane,
\begin{equation}
\label{eq:3_sided_map}
    w = \sqrt{e^{4\pi z/L}-1}.
\end{equation}
The three points in the correlator is mapped onto $\infty,i,-i$ on the plane.
\begin{figure}
    \centering
    \includegraphics[width = 0.85\linewidth]{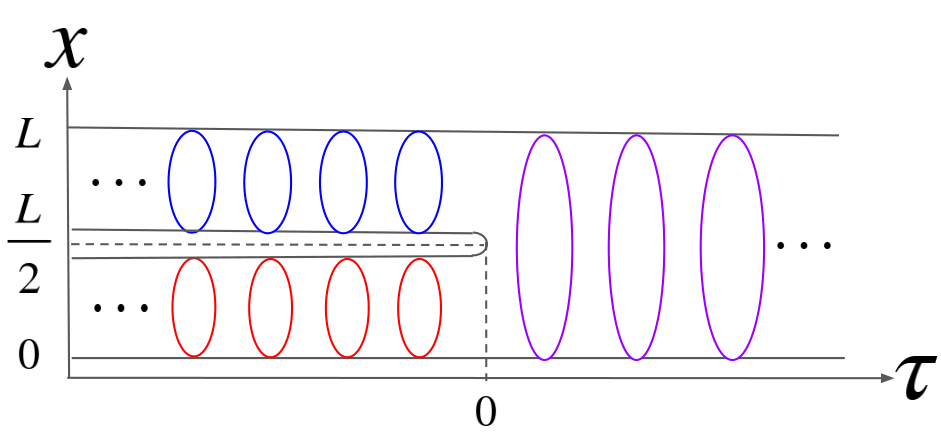}
    \includegraphics[width = 0.95\linewidth]{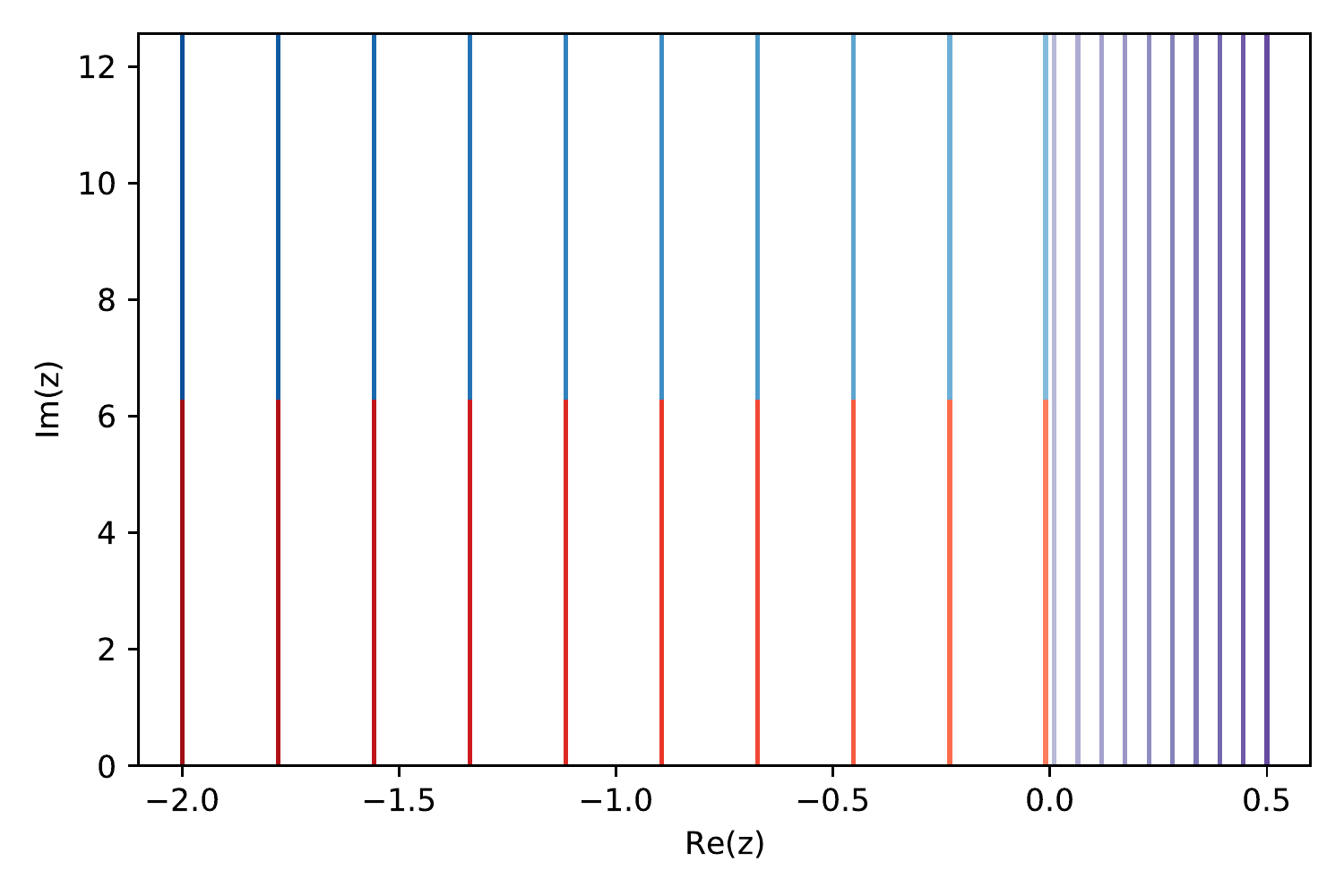}
    \includegraphics[width = 0.95\linewidth]{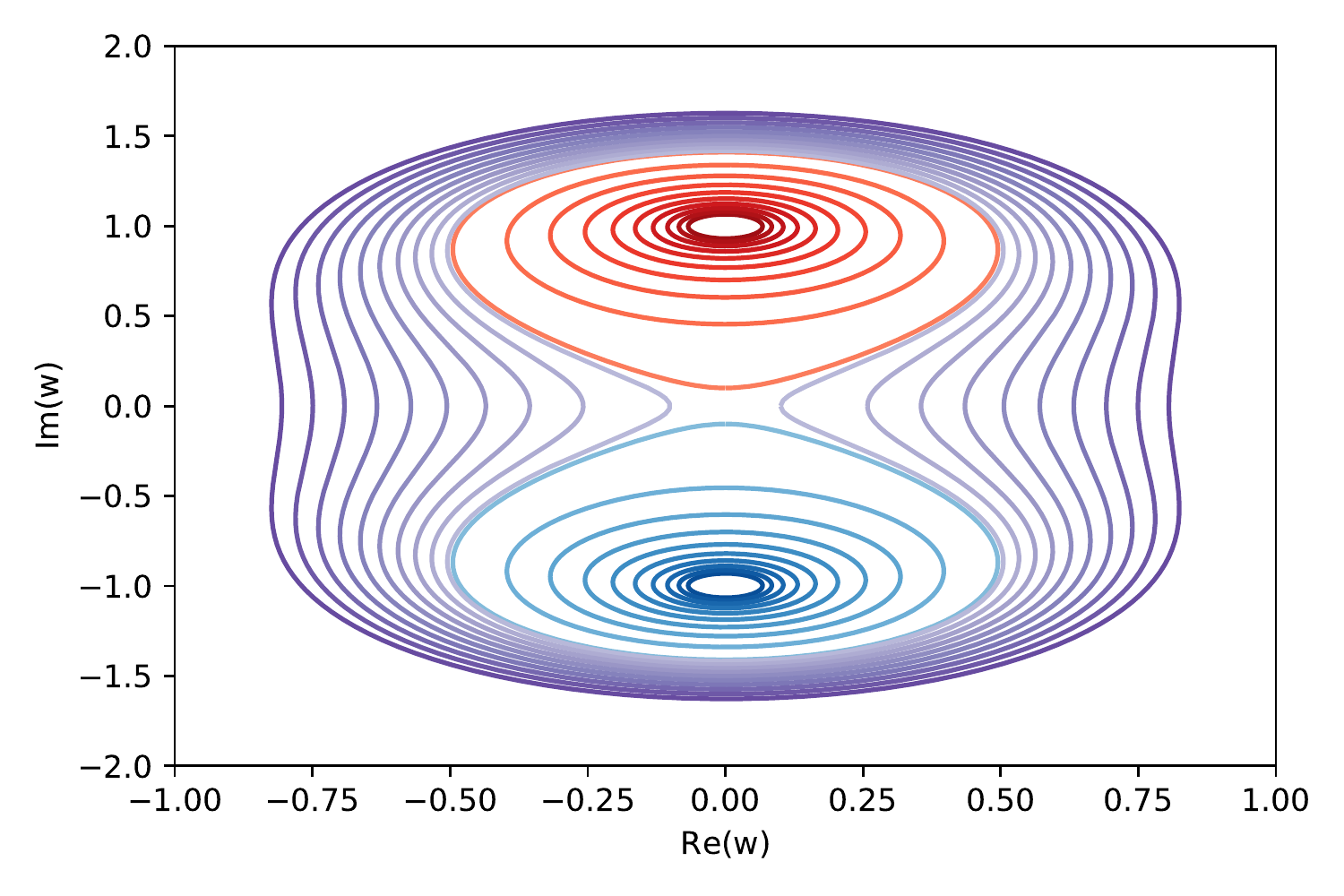}
    \caption{Mapping from three-sided wormhole to the complex plane. Top: An illustration of the pants geometry with different equal-time slices. Different colors label different regions of the geometry. Middle: Same as the top figure but we use straight lines to denote the equal-time slices. Different lightness labels different time slices, where darker colors label larger $|\tau|$. The circumference of the larger cylinder is taken to be $L=4\pi$.  Bottom: The image of the equal-time slices with the conformal mapping Eq.~\eqref{eq:3_sided_map}. The past infinity ($\tau=-\infty$) on two cylinders ($0<x<2\pi$ and $2\pi<x<4\pi$) and the future infinity $\tau=\infty$ are mapped to $i,-i$ and $\infty$, respectively.}
    \label{fig:conf_map}
\end{figure}

\begin{figure}
    \centering
    \includegraphics[width = 0.85\linewidth]{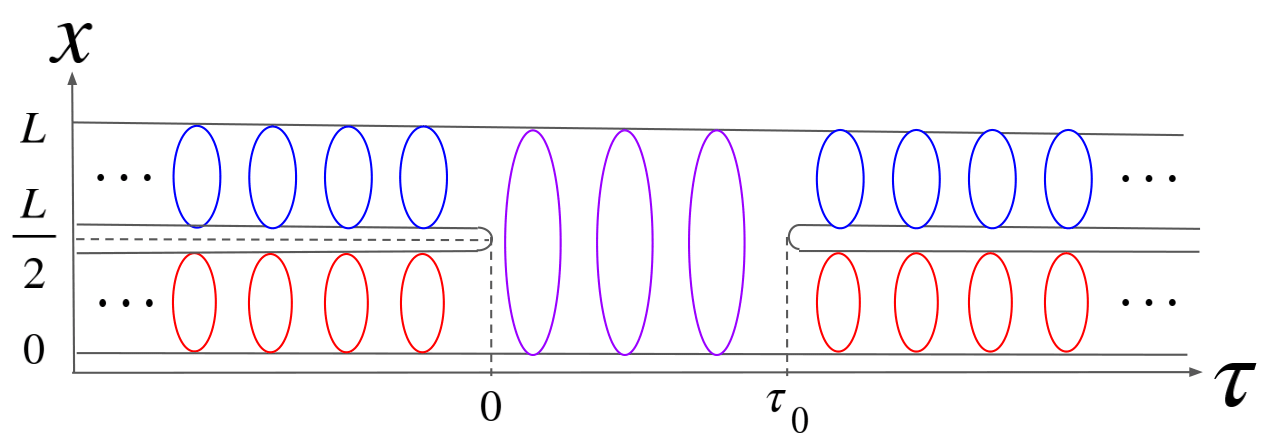}
    \includegraphics[width = 0.95\linewidth]{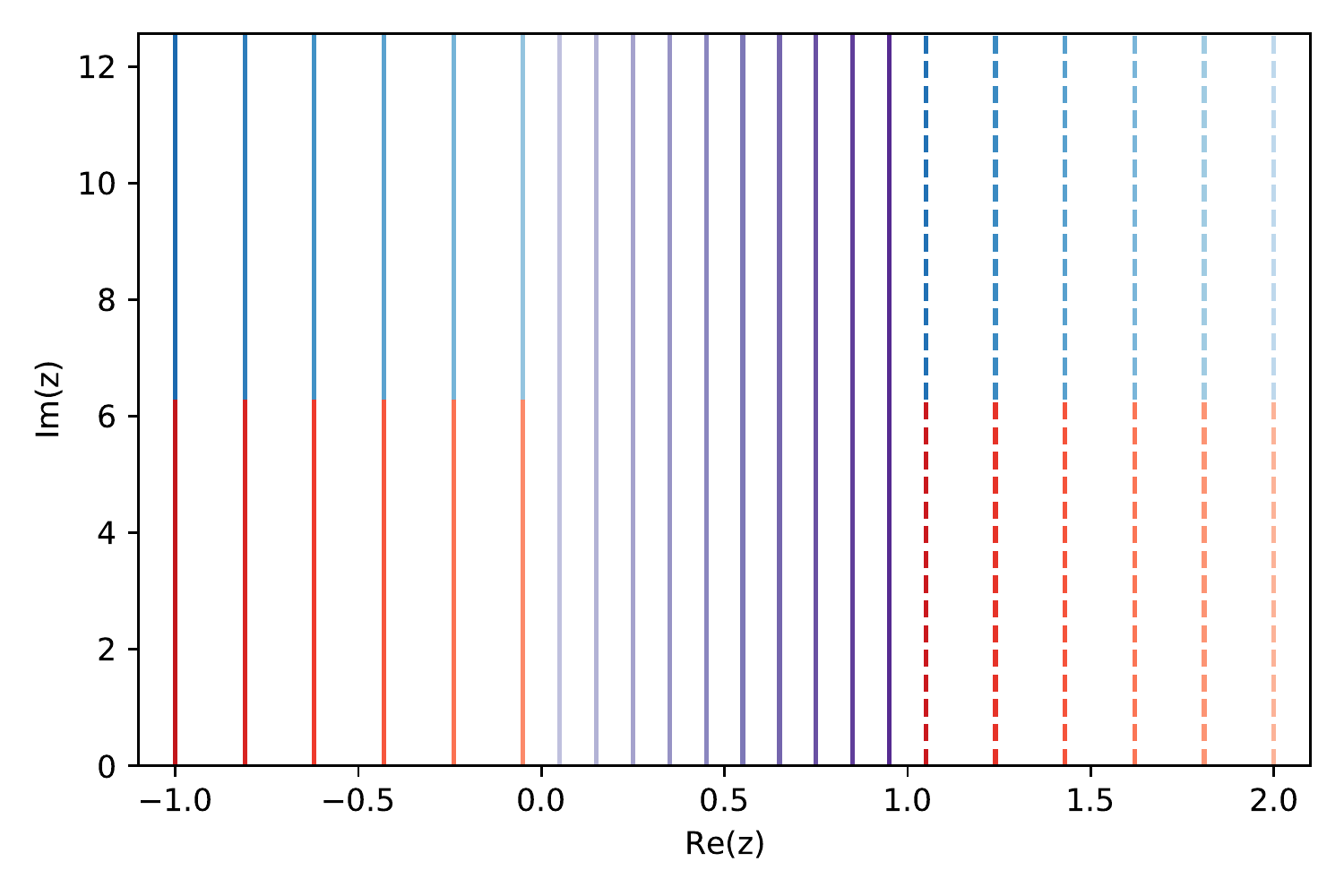}
    \includegraphics[width = 0.95\linewidth]{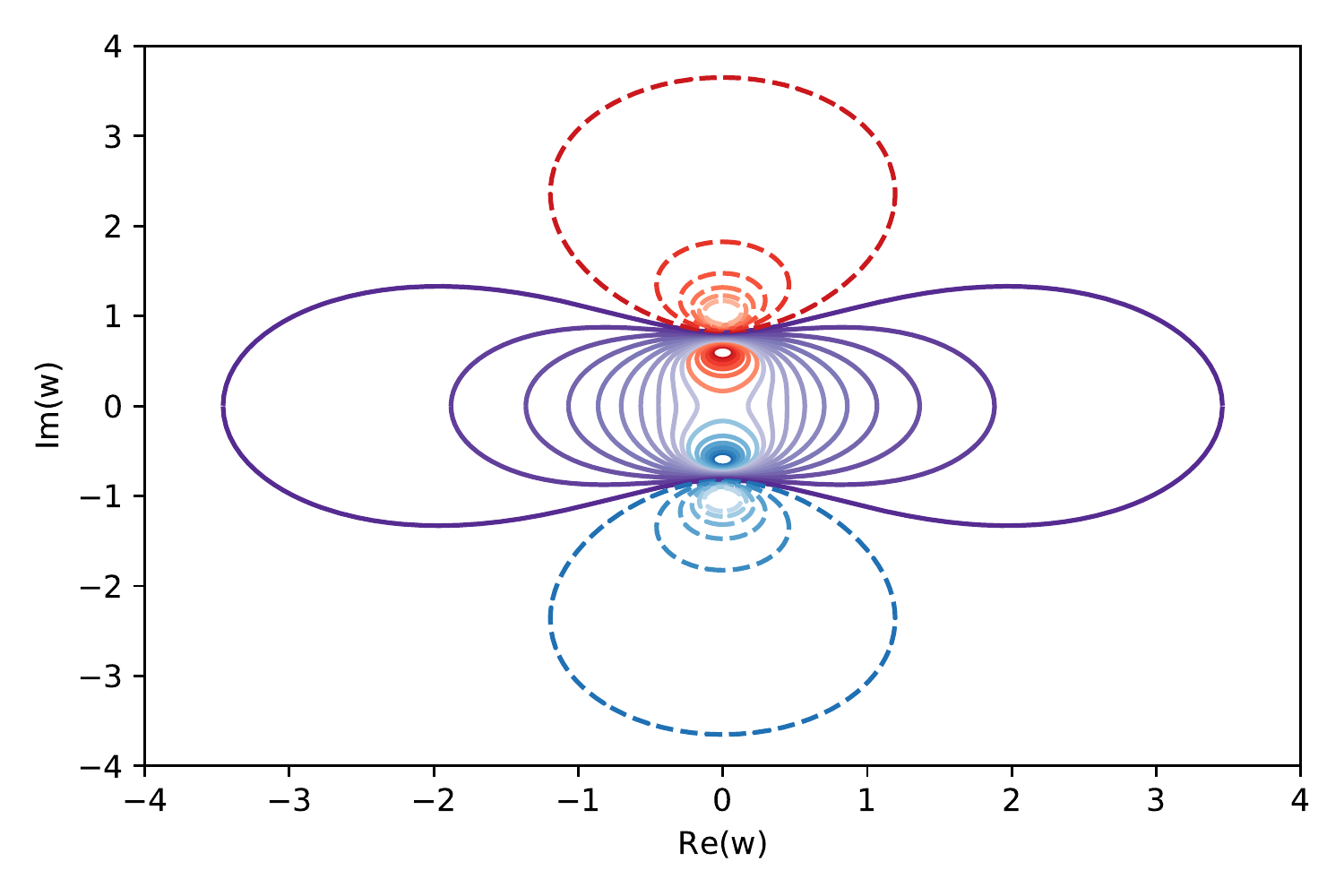}
    \caption{Mapping from four-sided wormhole geometry to the complex plane. Top: An illustration of the four-sided geometry with different equal-time slices. Different colors label different regions of the geometry. Middle: Same as the top figure but we use straight lines to denote the equal-time slices. Different lightness labels different imaginary time. The parameters are set to be $L=4\pi$ and $\tau_0=1$. Bottom: The image of the equal-time slice with the conformal mapping Eq.~\eqref{eq:4_sided_map}. The past infinity ($\tau=-\infty$) on two cylinders on the left ($0<x<2\pi$ and $2\pi<x<4\pi$) and the future infinity are mapped to $\pm ie^{-1/2}$. The future infinity on two cylinders on the right are mapped to $\pm i$.}
    \label{fig:conf_map4}
\end{figure}
\begin{equation}
    \frac{A_{\alpha\beta\gamma}}{A_{III}} = |J_1|^{\Delta_1} |J_2|^{\Delta_2} |J_3|^{\Delta_3} \langle\phi_{\alpha}(\infty)\phi_{\beta}(i)\phi_{\gamma}(-i)\rangle_{pl},
\end{equation}
where the $J$'s are Jacobian factors,
\begin{eqnarray}
    J_1 = \left(\frac{2\pi}{L}\right)^{-1}\lim_{\tau_1\rightarrow +\infty}w'(z_1) e^{-2\pi \tau_1/L} \\
    J_2 = \left(\frac{4\pi}{L}\right)^{-1}\lim_{\tau_2\rightarrow -\infty}w'(z_2) e^{-4\pi \tau_2/L}\\
    J_3 = \left(\frac{4\pi}{L}\right)^{-1}\lim_{\tau_3\rightarrow -\infty}w'(z_3) e^{-4\pi \tau_3/L}.
\end{eqnarray}
Working out the limits we obtain
\begin{eqnarray}
    J_1 = 1, ~~ J_2 =-i/2,~~ J_3 =i/2.
\end{eqnarray}
The three-point correlator is standard,
\begin{equation}
    \langle\phi_{\alpha}(\infty)\phi_{\beta}(i)\phi_{\gamma}(-i)\rangle_{pl} = 2^{\Delta_{\alpha}-\Delta_{\beta}-\Delta_{\gamma}}C_{\alpha\beta\gamma}.
\end{equation}
Finally we obtain
\begin{equation}
    \frac{A_{\alpha\beta\gamma}}{A_{III}} = 2^{-2\Delta_\beta-2\Delta_{\gamma}+\Delta_{\alpha}} C_{\alpha\beta\gamma}.
\end{equation}
\subsection{Overlap involving four states}
The four-sided TFD state $|\Omega_{12,34}(\tau_0)\rangle$ has wavefunction
\begin{equation}
    A_{\alpha\beta\gamma\delta}(\tau_0) = \langle \phi^{3}_{\gamma} \phi^{4}_{\delta}|e^{-\tau_0 H_0}|\phi^{1}_{\alpha} \phi^{2}_{\beta}\rangle. 
\end{equation}
There are two parameters $\tau_0$ and $L_0$ that determines the geometry. The geometry can be mapped onto the complex plane with the conformal transformation 
\begin{equation}
\label{eq:4_sided_map}
    w(z) = g(f(z)),
\end{equation}
where
\begin{eqnarray}
    f(z) = e^{4\pi z/L_0}  \\
    g(z) = \sqrt{\frac{z-1}{z_0-z}}
\end{eqnarray}
and 
\begin{equation}
    z_0 = e^{4\pi \tau/L_0}.
\end{equation}
Note that $f(z)$ maps the original geometry to a two-sheeted complex plane with branch cut on the $1<z<z_0$ segement of the real axis and $g(z)$ maps the latter into the complex plane. 
\begin{eqnarray}
\label{eq:ws}
    w(z_1) = i, w(z_2) = -i, \\
    \label{eq:ws2}
    w(z_3) = iz^{-1/2}_0, w(z_4) = -i z^{-1/2}_0
\end{eqnarray}
The overlap is mapped onto the four-point correlation function
\begin{equation}
\label{eq:CFT4pt}
    \frac{A_{\alpha\beta\gamma\delta}}{A_{IIII}} = \prod_{i} |J_i|^{\Delta_i} \langle \phi_{\alpha}(w_1) \phi_{\beta}(w_2) \phi_{\gamma}(w_3) \phi_{\delta}(w_4) \rangle_{pl},
\end{equation}
where the $J$'s are Jacobian factors,
\begin{eqnarray}
    J_1 &=& \left(\frac{4\pi}{L}\right)^{-1}\lim_{\tau_1\rightarrow +\infty}w'(z_1) e^{+4\pi (\tau_1-\tau_0)/L_0} \\
    J_2 &=& \left(\frac{4\pi}{L}\right)^{-1}\lim_{\tau_2\rightarrow +\infty}w'(z_2) e^{+4\pi (\tau_2-\tau_0)/L_0} \\
    J_3 &=& \left(\frac{4\pi}{L}\right)^{-1}\lim_{\tau_3\rightarrow -\infty}w'(z_3) e^{-4\pi \tau_3/L_0}\\
    J_4 &=& \left(\frac{4\pi}{L}\right)^{-1}\lim_{\tau_4\rightarrow -\infty}w'(z_4) e^{-4\pi \tau_4/L_0}.
\end{eqnarray}
A little bit algebra gives
\begin{eqnarray}
    J_1=-J_2=-\frac{i(z_0-1)}{2z_0} ,  \\
    J_3=-J_4=-i\frac{z_0-1}{2z^{3/2}_0}
\end{eqnarray}

\subsection{Ising CFT}
It is well known that four-point correlation functions in a CFT only depend on the a function of the cross ratio. For the case of four identical fields denoted as $\sigma$, the correlation function can be computed by 
\begin{eqnarray}
    &~&\langle \sigma(w_1) \sigma(w_2) \sigma(w_3) \sigma (w_4)\rangle  \nonumber \\
    &=& |w_3-w_1|^{-2\Delta_{\sigma}}|w_4-w_2|^{-2\Delta_{\sigma}}F_{\sigma\sigma\sigma\sigma}(\eta,\bar{\eta}),
\end{eqnarray}
where
\begin{equation}
    F_{\sigma\sigma\sigma\sigma}(\eta,\bar{\eta}) \equiv \langle \sigma(0)\sigma(\eta,\bar{\eta})\sigma(1) \sigma(\infty)\rangle,
\end{equation}
and $\eta$ is known as the cross ratio. With the $w$'s in Eqs.~\eqref{eq:ws} and \eqref{eq:ws2},
\begin{equation}
    \eta = tanh^2\left(\frac{\pi \tau_0}{L_0}\right).
\end{equation}
For the Ising CFT, it is exactly known that 
\begin{equation}
    F_{\sigma\sigma\sigma\sigma}(\eta,\bar{\eta}) = \frac{1}{2|\eta (1-\eta)|^{1/4}}(|1+\sqrt{1-\eta}| + |1-\sqrt{1-\eta}|),
\end{equation}
where $\sigma$ is the spin field. If $0\leq \eta \leq 1$, the formula simplifies to
\begin{equation}
    F_{\sigma\sigma\sigma\sigma}(\eta,\bar{\eta}) = (\eta (1-\eta))^{-1/4}.
\end{equation}
Substitution into Eq.~\eqref{eq:CFT4pt} yields
\begin{equation}
    \frac{A_{\sigma\sigma\sigma\sigma}}{A_{IIII}} = \frac{1}{2}\sqrt{\frac{1-e^{-4\pi\tau/L}}{tanh (\pi \tau/L)}}
\end{equation}
\section{Free fermion techniques}
Here we derive the new results of free fermion techniques used in the paper.
\subsection{Optimizing over smoothers}
The smoothers $U$ are optimized with the cost function of the form
\begin{equation}
    f(U) = |\langle \psi_T| U |\psi_0 \rangle|^2,
\end{equation}
where $|\psi_0\rangle$ and $|\psi_T\rangle$ are pure states with correlation matrices $M_0$ and $M_T$. In order to obtain the optimal $U$, we use a gradient descent optimization. At the first step, we let $U_0 = I$ and we update $U$ by 
\begin{equation}
    U = e^{\sum_{ij} \gamma_{i} \delta h_{ij} \gamma_j } U_0,
\end{equation}
where $\delta h$ is a real antisymmetric matrix with a small norm. The cost function to the first order is
\begin{equation}
    f(U) = |\langle \psi_T |\psi_0 \rangle|^2 + \delta h_{ij} G_{ij},
\end{equation}
where 
\begin{equation}
    G_{ij} = \langle \psi_T |\gamma_i \gamma_j|\psi_0\rangle \langle \psi_0|\psi_T\rangle + c.c.
\end{equation}
Now we use the following result from Ref.~\cite{Bravyi_2017}: 
\begin{equation}
    \frac{\langle \psi_T |\gamma_i \gamma_j |\psi_0\rangle}{\langle\psi_T|\psi_0\rangle} = i\Delta_{ij},
\end{equation}
where 
\begin{equation}
    \Delta = C_T - (I+iC_T)(C_0+C_T)^{-1}(I-iC_T).
\end{equation}
Then
\begin{equation}
    G_{ij} = |\Pf(C_0+C_T)|  \times i (\Delta_{ij} - \Delta^{*}_{ij})
\end{equation}
In the gradient descent method we choose
\begin{equation}
    \delta h_{ij} = \delta t G_{ij},
\end{equation}
where $\delta t$ is a tuning parameter that makes $f(U)$ largest. In the next step we substitute $C_0$ with
\begin{equation}
    C_0 \rightarrow e^{\delta h}C_0 e^{-\delta h}
\end{equation} 
and repeat. The norm of gradient is $||G||$, and the optimization error is on the order of $||G||^2$. We stop the optimization at $||G||<10^{-3}$, where $U$ is already sufficiently close to the optimum.

\subsection{Imaginary time evolution}
Let $H$ be a quadratic Hamiltonian 
\begin{equation}
    H = -\frac{i}{4}\sum_{ij} \gamma_i h_{ij} \gamma_j
\end{equation}
and $|\psi_0\rangle$ be a pure state with correlation matrix $M_0$. The imaginary time evolution
\begin{equation}
    |\psi(\tau)\rangle = \frac{1}{\sqrt{Z(\tau)}}e^{-\tau H}|\psi_0\rangle
\end{equation}
also reproduces a Gaussian state. The normalization constant equals
\begin{equation}
    Z(\tau) = \langle \psi_0| e^{-2\tau H} |\psi_0\rangle.
\end{equation}
To compute this, first we schur decompose the Hamiltonian matrix
\begin{equation}
    h = O (\diag(\epsilon_i)\otimes i\sigma^y) O^{T}
\end{equation}
Let
\begin{equation}
    \rho_{th} = e^{-2\tau H}/Z_{th},
\end{equation}
be the thermal state at inverse temperature $2\tau$, then
\begin{equation}
    Z_{th} = 2^N \prod_{i} cosh (\tau \epsilon_i)
\end{equation}
Let $\rho_0 = |\psi_0\rangle\langle\psi_0|$, the normalization constant can be computed by
\begin{equation}
    Z(\tau) = Z_{th}\Tr(\rho_0 \rho_{th})
\end{equation}
and 
\begin{equation}
    \Tr(\rho_0 \rho_{th}) = 2^{-N} |\Pf(M_0+M_{th})|,
\end{equation}
where 
\begin{equation}
    M_{th} = O (\diag(-tanh{\tau\epsilon_i})\otimes i\sigma^y) O^{T}
\end{equation}
is the correlation matrix of the thermal state. Combining together the above we obtain
\begin{equation}
    Z(\tau) =\prod_{i} \cosh(\tau\epsilon_i) |\Pf(M+M_{th})|.
\end{equation}
Next, we derive the correlation matrix of the output state
\begin{equation}
    M(\tau)_{ij} = \frac{1}{Z(\tau)}\Tr(-i (\gamma_i \gamma_j-\delta_{ij})e^{-\tau H} \rho_0 e^{-\tau H}) 
\end{equation}
Using the cyclic property of the trace, we obtain
\begin{equation}
    M(\tau)_{ij} = \frac{\Tr(-i (e^{-\tau H}\gamma_i \gamma_j e^{\tau H}-\delta_{ij})  \rho_{th} \rho_0)}{\Tr(\rho_{th}\rho_0)}
\end{equation}
Let $\Delta$ be 
\begin{equation}
    \Delta_{ij} = \frac{\Tr(-i (\gamma_i \gamma_j -\delta_{ij})  \rho_{th} \rho_0)}{\Tr(\rho_{th}\rho_0)},
\end{equation}
which equals \cite{Bravyi_2017}
\begin{equation}
    \Delta = M_{th} - (I+iM_{th})(M_0+M_{th})^{-1}(I-iM_{th}),
\end{equation}
then
\begin{equation}
    M(\tau) = e^{-i\tau h} \Delta e^{i\tau h}.
\end{equation}
Using the factor that $M^2_0=-I$ and with a bit matrix manipulations, we obtain
\begin{equation}
\label{eq:imag_time_evol}
    M(\tau) = M_{th} - \sqrt{I+M^2_{th}} (M_0+M_{th})^{-1} \sqrt{I+M^2_{th}}.
\end{equation}
\subsection{Canonical purification}
Given a Hamiltonian $H$ and an antiunitary operator $T$, the thermofield double can be defined by
\begin{equation}
\label{eq:TFD}
    |\beta\rangle = e^{-\frac{\beta}{4} (H_L+H_R)} |\Omega\rangle,
\end{equation}
where $H_L = H \otimes I$  and $H_R = I \otimes T H T^{-1}$, $|Omega\rangle$ is the maximal entangled state in the doubled Hilbert space. In our convention, we choose the antiunitary operator $T$ such that
\begin{equation}
    T \gamma_i T^{-1} = \gamma_i.
\end{equation}
This implies that
\begin{equation}
    T H T^{-1} = -H.
\end{equation}
The maximal entangled state is the simultaneous eigenstate of $-i \gamma_{iL} \gamma_{iR}$ with eigenvalue $1$. The correlation matrix us
\begin{equation}
    M_{\Omega} = \begin{bmatrix}
    0 & I  \\
    -I & 0
    \end{bmatrix}
\end{equation}
Applying the imaginary time evolution as in Eq.~\eqref{eq:TFD} and Eq.~\eqref{eq:imag_time_evol} we obtain the correlation matrix of $|\beta\rangle$,
\begin{equation}
\label{eq:M_TFD}
    M_{\beta} = \begin{bmatrix}
    M & \tilde{M} \\
    -\tilde{M} & -M
    \end{bmatrix}
\end{equation}
where 
\begin{eqnarray}
\label{eq: M_thermal}
    M &=& O (\diag(n_i)\otimes i\sigma^y) O^{T} \\
    \tilde{M} &=& O (\diag(\sqrt{1-n^2_i})\otimes I) O^{T} \\
    n_i &=& -tanh(\beta \epsilon_i/2).
\end{eqnarray}
Note that the reduced density matrix of the left/right subsystem is a thermal state with inverse temperature $\beta$ and Hamiltonian $H_L$/$H_R$.

Given a Gaussian reduced density matrix $\rho$ with correlation matrix $M$, we may define the entanglement Hamiltonian by 
\begin{equation}
    \rho = e^{-H_e}.
\end{equation}
The canonical purification is defined by
\begin{equation}
    |\sqrt{\rho}\rangle = e^{-\frac{1}{4}(H_{eL}+H_{eR})}|\Omega\rangle,
\end{equation}
which is the same as Eq.~\eqref{eq:TFD} except that $H$ is substituted by the entanglement Hamiltonian $H_e$. Furthermore, the entanglement Hamiltonian of a Guassian state is always quadratic. Thus, repeating the same calculation that leads to Eq.~\eqref{eq:M_TFD} we obtain the same formula, except that the $O$ and $n_i$ in Eq.~\eqref{eq: M_thermal} come from the Schur decomposition of $M$ rather than the Hamiltonian $h$.

Finally, in order to compute the reflected entropy, one starts with the correlation matrix $M_{\sqrt{\rho}}$ of the canonical purification and compute the entanglement entropy of the given subsystem.

\end{document}